\documentclass[amsmath,amssymb,aps,preprintnumbers,prd,twocolumn,superscriptaddress,showpacs]{revtex4}

\usepackage{times}
\usepackage{color}
\usepackage[normalem]{ulem}  
\usepackage{graphicx}
\usepackage{multirow}
\usepackage[colorlinks=true, pdfstartview=FitV, linkcolor=red, citecolor=blue, urlcolor=blue]{hyperref}
\renewcommand\sout{\bgroup \color{red} \ULdepth=-.5ex \ULset}

\newcommand{\Comments}[1]{}
\newcommand{\nn}{\nonumber}
\newcommand{\sub}[1]{{\scriptscriptstyle \mathrm{#1}}}

\newcommand{\Vp}{V^{+}}
\newcommand{\Vm}{V^{-}}

\newcommand{\Zp}{Z_{+}}
\newcommand{\Zm}{Z_{-}}

\newcommand{\bsigp}{b'_\sigma}
\newcommand{\bt}{\beta_\tau}

\newcommand{\btt}{\beta_{\tau\tau}}

\newcommand{\tilmu}{\tilde{\mu}}

\newcommand{\chibar}{{\bar{\chi}}}

\newcommand{\psiT}{\psi_{\tau}}
\newcommand{\psiS}{\psi_s}
\newcommand{\psiTT}{\psi_{\tau\tau}}
\newcommand{\psiTS}{\psi_{\tau s}}
\newcommand{\psiSS}{\psi_{ss}}

\newcommand{\psibarT}{\bar{\psi}_{\tau}}
\newcommand{\psibarS}{\bar{\psi}_s}
\newcommand{\psibarTT}{\bar{\psi}_{\tau\tau}}
\newcommand{\psibarTS}{\bar{\psi}_{\tau s}}
\newcommand{\lbar}{\bar{\ell}}
\newcommand{\psibarSS}{\bar{\psi}_{ss}}

\begin{document}
\preprint{YITP-16-119}

\title{Polyakov loop effects on the phase diagram in strong-coupling lattice QCD}

\author{Kohtaroh Miura}
\email[]{kohtaroh.miura@cpt.univ-mrs.fr}
\email[]{miura@kmi.nagoya-u.ac.jp}
\affiliation{Centre de Physique Theorique(CPT), Aix-Marseille University, \\
Campus de Luminy, Case 907, 163 Avenue de Luminy, 13288 Marseille cedex 9, France}
\affiliation{Kobayashi-Maskawa Institute for the Origin of Particles and the Universe, \\
Nagoya University, Nagoya 464-8602, Japan}
\author{Noboru Kawamoto}
\affiliation{Department of Physics, Faculty of Science, Hokkaido University, Sapporo 060-0810, Japan}
\author{Takashi Z. Nakano}
\affiliation{KOZO KEIKAKU ENGINEERING Inc., Tokyo 164-0012, Japan}
\author{Akira Ohnishi}
\affiliation{Yukawa Institute for Theoretical Physics, Kyoto University, Kyoto 606-8502, Japan}

\date{\today}

\begin{abstract}
We investigate the Polyakov loop effects on the QCD phase diagram
by using the strong-coupling $(1/g^2)$ expansion of the lattice QCD (SC-LQCD)
with one species of unrooted staggered quark, including $\mathcal{O}(1/g^4)$ effects.
We take account of the effects of Polyakov loop fluctuations in Weiss mean-field approximation (MFA),
and compare the results with those in
the Haar-measure MFA (no fluctuation from the mean-field).
The Polyakov loops strongly suppress the chiral transition temperature
in the second-order/crossover region at small chemical potential ($\mu$),
while they give a minor modification of the first-order phase boundary at larger $\mu$.
The Polyakov loops also account for a drastic increase of the interaction measure
near the chiral phase transition.
The chiral and Polyakov loop susceptibilities $(\chi_{\sigma},\chi_{\ell})$
have their peaks close to each other in the second-order/crossover region.
In particular in Weiss MFA, there is no indication of the separated
deconfinement transition boundary
from the chiral phase boundary at any $\mu$.
We discuss the interplay between the chiral and deconfinement dynamics via
the bare quark mass dependence of susceptibilities $\chi_{\sigma,\ell}$.
\end{abstract}
\maketitle

\section{Introduction}\label{sec:intro}
The phase diagram of quantum chromodynamics (QCD)
at finite temperature ($T$) and/or quark chemical potential ($\mu$)
\cite{Philipsen:2012nu,Fukushima:2013rx}
provides a deep insight into the Universe.
At the few microseconds after the big-bang,
a quark-gluon plasma (QGP)
is supposed to undergo the QCD phase transition/crossover,
which results in confinement of color degrees of freedom
and the dynamical mass generation of hadrons.
In fact, the first principle calculations
based on lattice QCD Monte Carlo simulations (LQCD-MC)
indicates the crossover around $T_c = 145 - 195$ (MeV) \cite{Borsanyi:2010bp}.
In compact star cores,
a cold-dense system would appear,
where various interesting phases are expected
\cite{CSC,McLerran:2007qj,QY-dev,McLerran:2009ve,chiral-spiral}.

The QCD phase transition can be investigated
in the laboratory experiments~\cite{Gale:2013da}:
Circumstantial experimental evidence
at the Relativistic Heavy-Ion Collider (RHIC)
in Brookhaven National Laboratory together with theoretical arguments
implies that the QGP is created in heavy-ion collisions
at $\sqrt{s_{\scriptscriptstyle{NN}}}=200~\mathrm{GeV}$,
and recent experiments at the Large Hadron Collider (LHC) in CERN
give stronger evidence.
Probing the phase diagram at finite $\mu$,
in particular the critical point (CP) \cite{CEP}, 
is a central topic in the ongoing and future heavy-ion collision experiments
at the Facility for Antiproton and Ion Research (FAIR) at GSI,
the Nuclotron-based Ion Collider fAcility (NICA) at JINR,
and the beam energy scan program at RHIC \cite{Mohanty:2011nm}.
Unfortunately, the first principle studies by LQCD-MC loses
the robustness at finite $\mu$ due to the notorious sign problem
\cite{Philipsen:2012nu,Muroya:2003qs,Aarts:2013bla,Fujii:2013sra,Cristoforetti:2012su}.
Many interesting subjects,
for example,
the location of CP,
the equation of state (EOS) at high density,
are still under debate.

The QCD phase diagram may be characterized by two underlying dynamics,
the chiral and deconfinement transitions,
which are associated with the spontaneous breaking of the chiral symmetry in the chiral limit
and the $Z_{N_c}$ center symmetry of the color SU$(N_c)$ gauge group in the heavy quark mass limit, respectively.
The order parameter is the chiral condensate ($\sigma$)/Polyakov loop ($\ell$) for the chiral/deconfinement transition.
Although the $Z_{N_c}$ symmetry is explicitly broken by the quark sector (with a finite or vanishing mass),
the Polyakov loops are still important degrees of freedom to be responsible for
the thermal excitation of quarks near the chiral phase transition.
The interplay between the $\sigma$ and $\ell$ is under active scrutiny;
the LQCD-MC reports that
the chiral and Polyakov loop susceptibilities show their peaks at almost the same temperatures for $\mu = 0$,
and the separation of two dynamics is proposed at finite $\mu$ in several models~\cite{Fukushima:2013rx}.

We investigate the QCD phase diagram by using
the strong-coupling expansion in the lattice QCD (SC-LQCD),
which provides a lattice-based and well-suited framework for the chiral and deconfinement transitions
without a serious contamination by the sign problem.
The SC-LQCD has been successful
since the beginning of the lattice gauge theory
\cite{TextBook,Wilson:1974sk,Creutz-MC,Munster:1980iv},
and revisited after the QGP discovery at RHIC
as an instructive guide to the QCD phase diagram
\cite{Fukushima:2003vi,
Nishida:2003uj,Nishida:2003fb,
Azcoiti:2003eb,
Kawamoto:2005mq,NLO,Miura:2011kq,Nakano:2009bf,PNNLO-T,
Bringoltz:2006pz,
deForcrand:2009dh,MDP-NLO,deForcrand:2014tha,
SC-LQCD-MC}.
It is remarkable that a promising
phase diagram structure has been obtained even in the
strong-coupling limit ($\beta=2N_c/g^2\to\infty$)
with mean-field approximation (MFA) \cite{Fukushima:2003vi,Nishida:2003fb,Kawamoto:2005mq},
and exactly determined based on the Monomer-Dimer-Polymer (MDP) formulation~\cite{deForcrand:2009dh}
and the Auxiliary Field Monte Carlo simulation~\cite{SC-LQCD-MC}.
The MFA results are then shown to be capturing the essential feature of the exact phase diagram.

\begin{figure*}[ht]
\includegraphics[width=15.0cm]{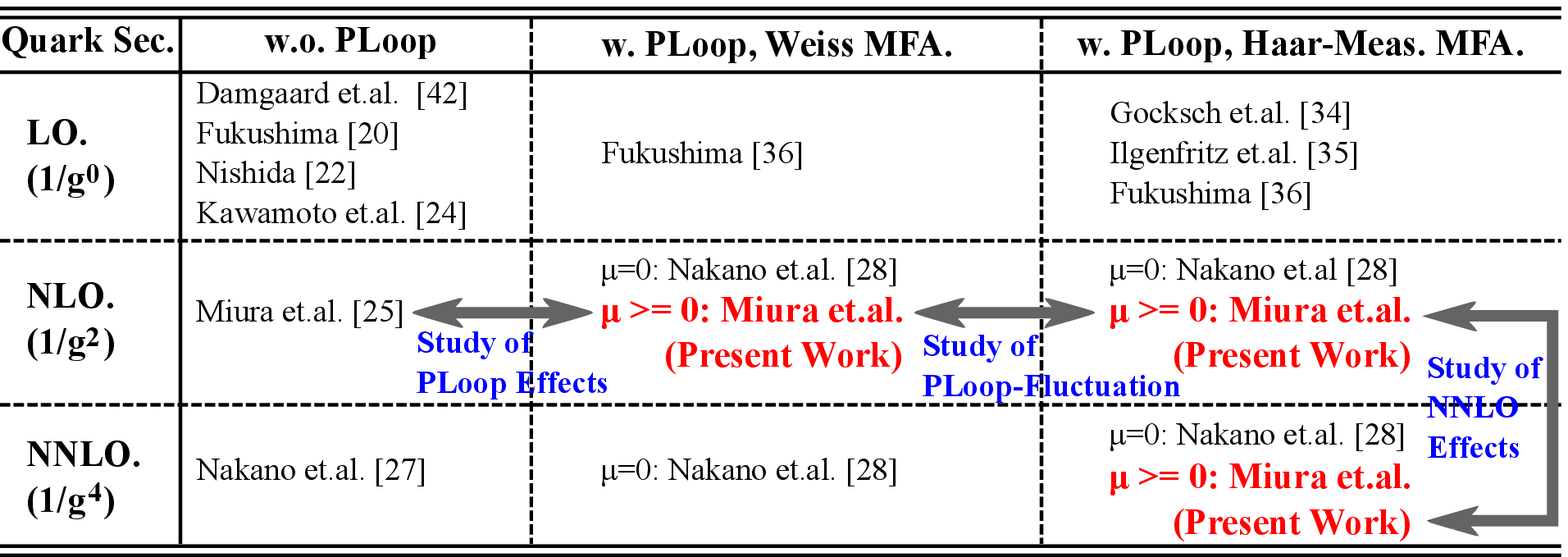}
\caption{(Color online)
The summary of the SC-LQCD studies on the QCD phase diagram for color SU(3) using MFA.
}\label{Fig:status}
\end{figure*}
In Fig.~\ref{Fig:status},
we summarize the SC-LQCD studies on the color SU(3) QCD phase diagram using MFA.
Based on the success in the strong-coupling limit (top in the second column),
we have investigated the phase diagram~\cite{NLO,Nakano:2009bf}
by taking account of the next-to-leading order (NLO, $\mathcal{O}(1/g^2)$, middle in the second column)
and the next-to-next-to-leading order (NNLO,$\mathcal{O}(1/g^4)$, bottom in the second column)
of the strong-coupling expansion.
The chiral phase transition temperature $T_c$ is strongly suppressed by the NLO effects,
and the phase diagram evolves into the empirical shape with increasing lattice coupling $\beta = 2N_c/g^2$,
while the NNLO effects give much milder corrections.

In the works mentioned above (listed in the second column of Fig.~\ref{Fig:status}),
the main focus was put on the chiral dynamics, rather than the $Z_{N_c}$ deconfinement dynamics,
which is another important dynamics described by the Polyakov loops $\ell$ of the pure-gluonic sector.
The SC-LQCD has been well-suited to include both dynamics
at the strong-coupling limit~\cite{Gocksch:1984yk,Ilgenfritz:1984ff,Fukushima:2003fm}
(top in third and fourth columns in Fig.~\ref{Fig:status});
the strong-coupling limit for the quark sector is combined with the leading-order effect
of the Polyakov loops in the pure-gluonic sector and the quark determinant term
provides the lattice-based derivation of the $\sigma$ - $\ell$ coupling.
It is intriguing to include the higher-order of the strong-coupling expansion,
which has been carried out in our previous work~\cite{PNNLO-T}
(middle and bottom lines in third and last columns in Fig.~\ref{Fig:status});
we have shown that the Polyakov loop effects combined with finite lattice couplings $\beta$
further suppresses the chiral transition temperature $T_c$,
which reproduces the results of LQCD-MC simulations \cite{Forcrand,MC4,Engels:1996ag} at $\mu = 0$
in the certain lattice coupling range $\beta\sim 4$.
Thus, the long-standing problem of the SC-LQCD - too large $T_c$ -
is greatly relaxed by the Polyakov loops.
Moreover, the Polyakov loop sector at the chiral phase transition $\sim \mathcal{O}([1/g^2]^{1/T_c})$
is found to be comparable with the quark sector with NLO [${\cal O}(1/g^2)$] and NNLO [${\cal O}(1/g^4)$]
at $T_c(\beta\sim 4) \sim 0.5 - 0.6$ (in lattice units);
the Polyakov loop effects are necessary to evaluate $T_c$
with respect to the order counting of the strong-coupling expansion.

In our previous paper~\cite{PNNLO-T}, however,
the analysis was limited at vanishing chemical potential $\mu = 0$,
while the finite $\mu$ region receives a growing interest by the forthcoming experiments
focusing the CP and high density phase.
The purpose of the present paper
is to extend our previous work~\cite{PNNLO-T} to the finite $\mu$ region,
and to investigate the Polyakov loop effects on the whole region of the QCD phase diagram
as indicated by red-solid characters in Fig.~\ref{Fig:status}.
We adopt two approximation schemes for the Polyakov loops,
a simple mean-field treatment (Haar-measure MFA)
and an improved treatment with fluctuation effects (Weiss MFA).
Through the various comparisons indicated by the arrows in Fig.~\ref{Fig:status},
we elucidate the effect of the Polyakov loop itself, either the effects of the Polyakov loop fluctuations,
as well as the higher-order (NNLO) effects of the strong-coupling expansion.
In particular, we focus on thermodynamic quantities,
which is of great interest in the study of the equation of state for quark matter
but has been challenging in SC-LQCD.
Moreover, we discuss the interplay between the chiral and deconfinement dynamics at finite $\mu$
via the bare quark mass dependence of susceptibilities $\chi_{\sigma,\ell}$.

We employ one species (unrooted) of staggered fermion,
which has a $U_{\chi}(1)$ chiral symmetry in the strong-coupling region
and becomes the four flavor QCD with degenerate masses in the continuum limit.
We investigate the $U_{\chi}(1)$ chiral phase transition/crossover at finite $T$ and $\mu$
in color SU($N_c=3$) gauge group in the $3+1$ dimension ($d=3$).
Our focus is not necessarily put on quantitative prediction of the realistic phase diagram,
but we attempt to clarify which effects make the SC-LQCD phase diagram being closer to realistic one.
Such lattice based arguments would be instructive
to future LQCD-MC studies on the QCD phase diagram,
even though the flavor-chiral structure in the present study
is different from the real-life QCD with 2+1 flavors.

This paper is organized as follows:
In Sec.~\ref{sec:Feff},
we explain the effective potential
in strong-coupling lattice QCD with Polyakov loop effects.
In Sec.~\ref{sec:result}, we investigate the phase diagram
and related quantities
by using the effective potential.
In Sec.~\ref{sec:summary}, we summarize
our work and give a future perspective.
Appendix \ref{app:Feff} is devoted to the review
of the effective potential derivation.

\section{Strong-coupling lattice QCD with Polyakov loop effects}
\label{sec:Feff}
We explain the effective potential
of the strong-coupling lattice QCD
including the Polyakov loop effects.
The derivation has been detailed in our previous work~\cite{PNNLO-T},
and recapitulated in Appendix \ref{app:Feff} in this paper.
Here we explain the essential property of the effective potential.
We will work on lattice units
$a = 1$ in color SU($N_c=3$) gauge and 3+1 dimension ($d=3$).
The parameters in the effective potential are
the lattice coupling $\beta=2N_c/g^2$,
lattice bare quark mass $m_0$,
lattice temperature $T=1/N_t$ ($N_t$ = temporal lattice extension),
and quark chemical potential $\mu$.

The effective potential $\mathcal{F}_{\mathrm{eff}}^{\sub{H/W}}$
involves
the plaquette-driven Polyakov loop sector $\mathcal{F}_{\sub{P}}^{\sub{H/W}}$ and
the quark sector $\mathcal{F}_{\sub{Q}}^{\sub{H/W}}$,
\begin{align}
&\mathcal{F}_{\mathrm{eff}}^{\sub{H/W}}(\Phi,\ell,\bar{\ell};\beta,m_0,T,\mu)\nn\\
&=\mathcal{F}_{\sub{P}}^{\sub{H/W}}(\ell,\bar{\ell},\beta,T)
+\mathcal{F}_{\sub{Q}}^{\sub{H/W}}(\Phi,\beta,m_0,T,\mu)\nn\\
&\quad +\mathcal{O}(1/g^6,1/g^{2(N_t+2)},1/\sqrt{d})
\ .\label{Eq:Feff}
\end{align}
The $\mathcal{F}_{\sub{P}}^{\sub{H/W}}$
is responsible for the Polyakov loop effects
\begin{align}
L_p = N_c^{-1}\prod_{\tau}U_{0,\tau\mathbf{x}}\ ,\quad
U_0 = \text{temporal link variable}\ ,
\end{align}
which result from the integral over the spatial link variables for
the plaquettes wrapping around the temporal direction.
Such Polyakov loops are dubbed ``plaquette-driven,'' and purely gluonic.
The effects of $L_p$ is investigated in two MFA scheme:
the Haar measure and Weiss MFA
as indicated by the suffixes ``H'' and ``W''.
In the former, the Polyakov loop $L_p$
is simply replaced with its constant mean-field $\ell$,
while in the latter,
the mean-field $\ell$ is introduced via the extended Hubbard-Stratonovich transformation \cite{NLO}
and the fluctuations from the mean-field
is taken account in the integral over the $U_0$.
The Polyakov loop effective potential 
of Haar measure MFA is well-known since the 1980s \cite{V-Ploop},
\begin{align}
&\mathcal{F}_{\sub{P}}^{\sub{H}}(\ell,\bar{\ell},\beta,T)\nn\\
&=-2TdN_c^2\biggl(\frac{1}{g^2N_c}\biggr)^{1/T}
\bar{\ell}\ell - T\log \mathcal{R}_{\mathrm{Haar}}\ ,\label{Eq:FPol_Haar} \\
&\mathcal{R}_{\mathrm{Haar}}
\equiv
1-6\bar{\ell}\ell
-3\bigl(\bar{\ell}\ell\bigr)^2
+4\bigl(\ell^{N_c}+\bar{\ell}^{N_c}\bigr)
\ ,\label{Eq:R_Haar}
\end{align}
where the Haar measure in the $U_0$ path integral
leads to the $Z_3$ symmetric term $\mathcal{R}_{\mathrm{Haar}}$.
Since the $\mathcal{R}_{\mathrm{Haar}}$ does not couple to the dynamical quarks,
the $Z_3$ symmetry affects the phase diagram separately from the chiral dynamics in Haar measure MFA.
In sharp contrast to this,
there is no counterpart in Weiss MFA \cite{PNNLO-T},
\begin{align}
&\mathcal{F}_{\sub{P}}^{\sub{W}}(\ell,\bar{\ell},\beta,T)
= 2TdN_c^2\biggl(\frac{1}{g^2N_c}\biggr)^{1/T} \bar{\ell}\ell
\ .\label{Eq:FPol_Weiss}
\end{align}
The Polyakov loop effects other than the quadratic term (\ref{Eq:FPol_Weiss})
are entangled to the dynamical quarks 
in the quark determinant as explained in the followings.
Thus, the $Z_3$ dynamics is totally spoiled by the dynamical quarks in Weiss MFA.

In both Haar measure and Weiss MFA cases,
the order counting of the strong-coupling expansion reads,
\begin{align}
&\mathcal{F}_{\sub{P}}^{\sub{H/W}}
\sim \mathcal{O}((1/g^2)^{N_t = 1/T})\ ,\label{Eq:gg_ploop_action}
\end{align}
and thus depends on the lattice temperature $T = 1/N_t$, which is subject to the integer value $N_t$.
However in this paper, we regard $T$ as a continuous valued given number,
which naturally follows in the lattice Matsubara formalism \cite{DKS}.
Around the chiral transition/crossover temperature $T_c$,
we will show that the $\mathcal{F}_{\sub{P}}^{\sub{H/W}}$ becomes comparable to
the NLO or NNLO effects: $\mathcal{O}(1/g^{2/T_c})\sim \mathcal{O}(1/g^{2 - 4})$.

The quark sector $\mathcal{F}_{\sub{Q}}^{\sub{H/W}}$ in Eq.~(\ref{Eq:Feff})
is derived by integrating out the staggered quarks with link/plaquette variables
in each order of the strong-coupling expansion.
In this paper, we consider the LO, NLO, and NNLO effects;
\begin{align}
&\mathcal{F}_{\sub{Q}}^{\sub{H/W}}
\ni \mathcal{O}(1/g^0)\ ,\mathcal{O}(1/g^2)\ ,\mathcal{O}(1/g^4)
\ .\label{Eq:gg_quark_action}
\end{align}
The integral is evaluated by introducing several auxiliary fields $\Phi$,
which includes the chiral condensate $\sigma$, the order parameter of the $U_{\chi}(1)$ chiral symmetry,
as well as other fields,
\begin{align}
\Phi=
\bigl\{
\sigma,
\psi_{\tau},
\bar{\psi}_{\tau},
\psi_s,
\bar{\psi}_s,
\psi_{\tau s},
\bar{\psi}_{\tau s},
\psi_{ss},
\bar{\psi}_{ss},
\psi_{\tau\tau},
\bar{\psi}_{\tau\tau}
\bigr\}\ ,
\end{align}
whose physical meanings are summarized in Tables~\ref{Tab:composites} and \ref{Tab:aux} in the Appendix \ref{app:Feff}.
The coefficients of the effective potential terms are solely characterized by ($\beta,N_c,d$)
and $\mathcal{O}(1/g^{0 - 4})$ (see Table \ref{Tab:coupling}).
The total quark sector $\mathcal{F}_{\sub{Q}}^{\sub{H/W}}$ is then divided into
the auxiliary field part
$\mathcal{F}_{\sub{X}}$
and the quark determinant part
$\mathcal{F}_{\mathrm{det}}^{\sub{H/W}}$.
As shown in Eq.~(\ref{Eq:Faux_app}) in Appendix \ref{app:Feff},
the $\mathcal{F}_{\sub{X}}$
is composed of the quadratic terms of the auxiliary fields $\Phi$.

The quark determinant term
$\mathcal{F}_{\mathrm{det}}^{\sub{H/W}}$ 
is responsible for the dynamical quark effects,
and includes the quark hoppings with
link variables $U_0$ wrapping around the temporal direction,
which give rise to the ``quark-driven'' Polyakov loops.
In Haar measure MFA, the quark determinant part becomes similar to that
in the Polyakov-loop-extended Nambu–-Jona-Lasinio (PNJL) model~\cite{Fukushima:2003fw,Fukushima:2008wg}
\footnote{In fact, the PNJL model is invented from the SC-LQCD~\cite{Fukushima:2003fw}.}
and the Polyakov-loop-extended Quark-Meson (PQM) model~\cite{Herbst:2010rf}:
\begin{align}
&\mathcal{F}_{\mathrm{det}}^{\sub{H}}
= -N_cE_q - N_c \log \sqrt{Z_+Z_-}\nn\\
&\quad
-T\Bigl(\log\mathcal{R}_{q}(E_q-\tilmu,\ell,\lbar)
+\log\mathcal{R}_q(E_q+\tilmu,\lbar, \ell)
\Bigr)\ ,\label{Eq:FDet_Haar}\\
&\mathcal{R}_{q}(x,y,\bar{y})
\equiv 1+N_c(y e^{-x/T} + \bar{y} e^{-2x/T})+e^{-3x/T}
\ .\label{Eq:Rq}
\end{align}
See Table \ref{Tab:renorm} for the quark excitation energy $E_q$,
the shifted quark chemical potential $\tilde{\mu}$,
and the wave function renormalization factor $\sqrt{Z_+Z_-}$.
In Weiss MFA, the plaquette-driven and quark-driven Polyakov loops are combined
in the quark determinant, and the $U_0$ path integral accounts for the Polyakov loop fluctuations.
Then we obtain the following expression,
\begin{align}
&\mathcal{F}_{\mathrm{det}}^{\sub{W}}
= - N_c \log \sqrt{Z_+Z_-}
-T\log\biggl[
\sum_{I}
\mathcal{Q}^{I}(\Phi)
\mathcal{P}^{I}(\ell,\bar{\ell})~
\biggr]\ ,\label{Eq:FDetW}\\
&\mathcal{P}^{I}(\ell,\bar{\ell})
=
\sum_{n=-\infty}^{\infty}
\biggl(
\sqrt{\ell/\bar{\ell}}
\biggr)^{-N_cn + N_{\mathrm{Q}}^I}
\mathcal{P}^{I}_n\biggl(\sqrt{\ell\bar{\ell}}\biggr)
\ ,\label{Eq:Pn}
\end{align}
where the thermal excitation of a quark and its composite
$\mathcal{Q}^{I}$,
the thermal excitation of Polyakov loops
$\mathcal{P}^{I}_n$,
and the quark number index $N_{\mathrm{Q}}^{I}$
are summarized in Table \ref{Tab:FDetW} in Appendix \ref{app:Feff}.
In the heavy quark limit $m_0\to \infty$,
Eq.~(\ref{Eq:FDetW}) recovers the $Z_3$ symmetry as shown in Appendix \ref{app:Feff}.

The auxiliary fields $\{\Phi,\ell,\bar{\ell}\}$ at equilibrium
are determined as a function of ($\beta,m_0,T,\mu$)
via the saddle point search of the effective potential
$\mathcal{F}^{\sub{H/W}}_{\mathrm{eff}}$.
In particular, the important quantities to probe the phase diagram
are the chiral condensate $\sigma\in \Phi$,
Polyakov loops ($\ell,\bar{\ell}$),
and their (dimensionless) susceptibilities
($\chi_{\sigma},\chi_{\ell}$).
In the present mean-field framework,
the susceptibilities are evaluated as follows:
We consider the curvature matrix $C$ of the effective potential at equilibrium,
\begin{align}
C_{ij}
= \frac{1}{T^4}
\frac{\partial^2 \mathcal{F}^{\sub{H/W}}_{\mathrm{eff}}}
{\partial \phi_i \partial \phi_j}\Big|_{\mathrm{equilibrium}}
\ ,\label{Eq:d2V}
\end{align}
where the field $\phi_i$ represents the dimensionless auxiliary fields
normalized by $T$ and $N_c$,
\begin{align}
&\phi_i \in
\biggl\{
\frac{\sigma}{T^3N_c},
\frac{\psi_{\tau}}{T^3N_c},
\frac{\bar{\psi}_{\tau}}{T^3N_c},
\frac{\psi_s}{T^6N_c^2},
\frac{\bar{\psi}_s}{T^6N_c^2},\nn\\
&\qquad
\frac{\psi_{\tau s}}{T^6N_c^2},
\frac{\bar{\psi}_{\tau s}}{T^6N_c^2},
\frac{\psi_{ss}}{T^{12}N_c^4},
\frac{\bar{\psi}_{ss}}{T^6N_c^2},
\frac{\psi_{\tau\tau}}{T^3N_c},
\frac{\bar{\psi}_{\tau\tau}}{T^3N_c},
\ell,\bar{\ell}
\biggr\}\ .
\end{align}
Then the chiral and Polyakov loop susceptibilities are given by
\begin{align}
\chi_{\sigma} = (C^{-1})_{ij=\sigma\sigma}\ ,\quad
\chi_{\ell} = (C^{-1})_{ij=\ell\bar{\ell}}\ .\label{Eq:sus}
\end{align}
In addition,
we investigate thermodynamic quantities,
a pressure $p$,
quark number density $\rho_q$,
and interaction measure $\Delta$,
\begin{align}
& p =
-\bigl(
\mathcal{F}_{\mathrm{eff}}^{\sub{H/W}}(T,\mu) - 
\mathcal{F}_{\mathrm{eff}}^{\sub{H/W}}(0,0)
\bigr)
\ ,\label{Eq:press}\\
&\rho_q =
\frac{\partial p}{\partial \mu}
\ ,\label{Eq:dens}\\
&\Delta = 
\frac{\epsilon - 3p}{T^4}
\ ,\label{Eq:i_meas}
\end{align}
where $\epsilon = -p + Ts  + \mu\rho_q$
represents an internal energy
with $s = \partial p/\partial T$ being an entropy.

\medskip

\section{Results}\label{sec:result}
We investigate the QCD phase diagram based on the effective potential
explained in the previous section.
We show the phase diagram and related quantities obtained in
Haar measure MFA at next-to-leading order (NLO) in Subsec. \ref{subsec:NLO_Haar},
Weiss MFA at NLO in Subsec.~\ref{subsec:NLO_Weiss} for the fixed lattice coupling $\beta = 4$
in the chiral limit ($m_0=0$).
We extend our study to include the finite bare quark mass $m_0 > 0$ in Subsec.~\ref{subsec:qmass}
with a particular focus on the chiral and Polyakov loop susceptibilities.
Then, in Subsec.~\ref{subsec:PD_evol}, we show the phase diagram evolution for various $\beta$.
Finally, in Subsec.~\ref{subsec:NNLO_Haar},
we study the next-to-next-to-leading order (NNLO) effects in the phase diagram.
The quark mass $m_0$, temperature $T$, quark chemical potential $\mu$, and other quantities
are all given in lattice units, unless explicitly stated otherwise.
\subsection{Haar measure MFA at NLO}\label{subsec:NLO_Haar}

We consider the NLO Haar measure MFA,
where the NNLO $\mathcal{O}(1/g^4)$ terms in the coupling coefficients
shown in Table \ref{Tab:coupling} and the Polyakov loop fluctuations are ignored.
We concentrate on the chiral limit case $m_0 = 0$.
We take the lattice coupling $\beta = 4.0$ as a typical value,
for which the chiral transition temperature at vanishing quark chemical potential
$T_{c,\mu = 0}$ \cite{PNNLO-T} becomes close to the LQCD-MC result~\cite{Forcrand}
(For details on the comparison, see Refs.~\cite{PNNLO-T,deForcrand:2014tha}).
The effects ignored or restricted here 
will be investigated in later subsections.
The phase diagram in the Haar measure MFA is partly studied
in our previous work~\cite{Miura:2011kq},
and we provide more complete analyses in the followings.

\begin{figure}[tbhp]
\includegraphics[width=8.0cm]{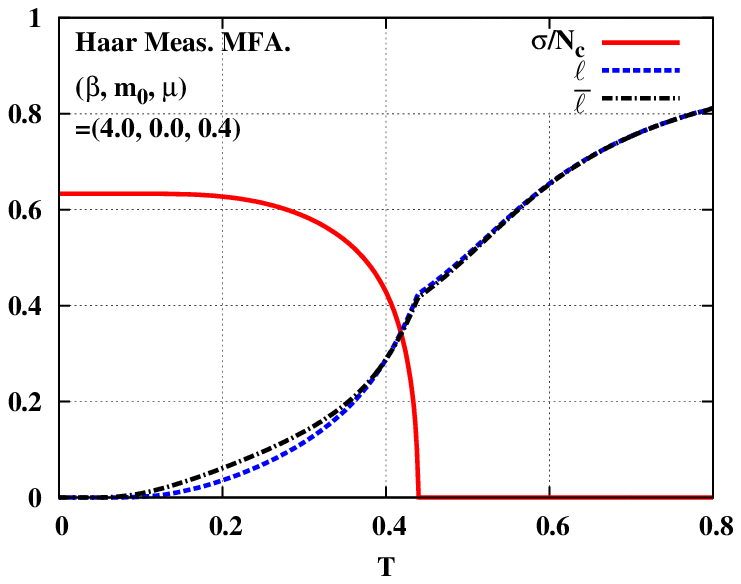}
\includegraphics[width=8.0cm]{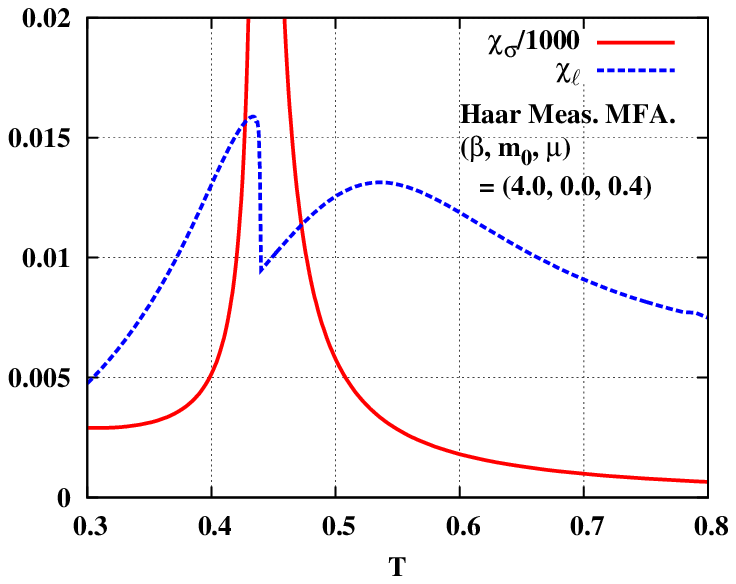}
\caption{(Color online) 
Upper:
The chiral condensates $\sigma$, Polyakov loops $(\ell,\bar{\ell})$,
in NLO Haar measure MFA as a function of $T$
at $(\beta,m_0,\mu) = (4.0,0.0,0.4)$ in lattice units.
Lower:
The chiral and Polyakov loop susceptibilities
($\chi_{\sigma}$ and $\chi_{\ell}$)
in the same condition as the upper panel in lattice units.
For a comparison, the $\chi_{\sigma}$ is multiplied by 1/1000.
}\label{Fig:H_40_00_T_04}
\end{figure}

In the upper panel of Fig.~\ref{Fig:H_40_00_T_04},
we show the chiral condensates ($\sigma/N_c$)
and Polyakov loops ($\ell,\bar{\ell}$)
at finite quark chemical potential $\mu = 0.4$
as a function of temperature $T$
for $(\beta,m_0) = (4.0,0.0)$.
In the low $T$ region, the chiral broken $(\sigma \neq 0)$
and confined $(\ell \sim 0)$ phase appears.
As $T$ increases,
we observe the second-order chiral phase transition $(\sigma \to 0)$
at $T_c\simeq 0.44$
and the large increase of the Polyakov loops
$(\ell \to \mathcal{O}(1))$.
These results are similar to the zero chemical potential case
shown in the previous study \cite{PNNLO-T}.

We find that the Polyakov loop is smaller
than the anti-Polyakov loop
($\ell < \bar{\ell}$) in the chiral broken phase.
This is understood from a quark screening effect at high density:
A finite $\mu$ leads to a net quark number density at equilibrium,
where putting additional quarks into the system would
give a larger energy cost than antiquarks.
Therefore the free energy of the quark
gets larger than that of the antiquark $F_{q}> F_{\bar{q}}$,
which attributes to our observation $\ell < \bar{\ell}$
through the relation
$(\ell,\bar{\ell}) \propto (e^{-F_q/T},e^{-F_{\bar{q}}/T})$.

In the lower panel of Fig.~\ref{Fig:H_40_00_T_04},
we compare the temperature dependence
of the chiral and Polyakov loop susceptibilities ($\chi_{\sigma},\chi_{\ell}$)
which are defined in Eq.~(\ref{Eq:sus})
in the same condition as the upper panel.
The Polyakov loop susceptibility has two peaks
with a relatively wide width.
We note that the action in the present SC-LQCD (\ref{Eq:Z}) 
has the $U_{\chi}(1)$ chiral symmetry,
which governs the dynamics of the system.
Since the first peak is found
in the vicinity of the chiral phase transition,
it should be associated with the chiral dynamics.
For example, the $\chi_{\ell}$ rapidly (but continuously) decreases
just after the peak, and its derivative with respect to $T$ is discontinuous.
This property is associated by the second-order chiral phase transition,
\begin{align}
\sigma (T) \propto
\begin{cases}
\left(\frac{T_c - T}{T_c}\right)^{\beta_\sigma = 1/2} & (T < T_c)\ ,\\
0            & (T \geq T_c)\ ,
\end{cases}
\label{Eq:sig_Tc}
\end{align}
via the potential curvature matrix (13).
In general, fluctuation effects modifies the critical exponent $\beta_\sigma$,
but the derivative is still discontinuous in the thermodynamic limit
at the second-order transition.
The second peak (or bump) is found
in the chiral restored phase $T\simeq 0.53 > T_c$,
and interpreted as the remnant
of the $Z_3$ deconfinement dynamics
as discussed in the subsection \ref{subsec:qmass}.

\begin{figure}[tbhp]
\includegraphics[width=8.0cm]{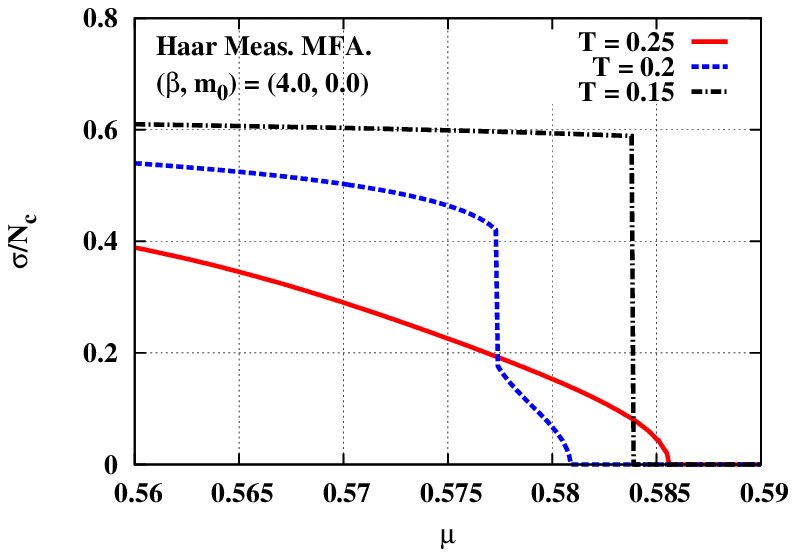}
\includegraphics[width=8.0cm]{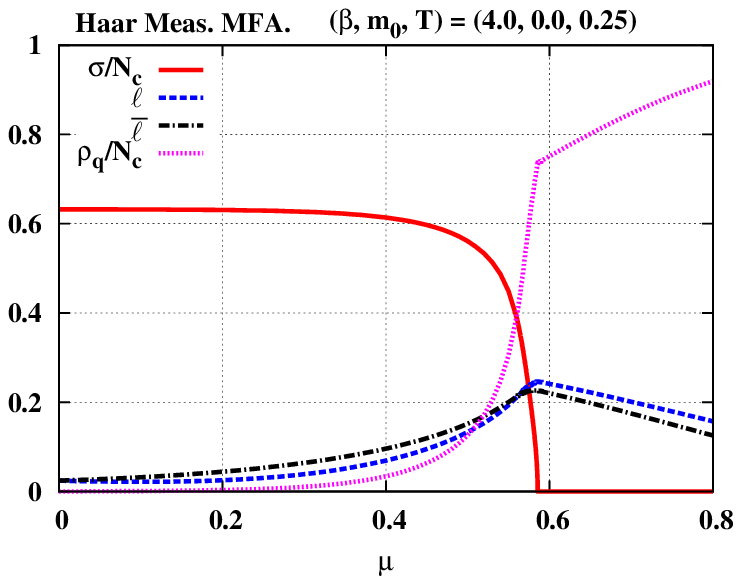}
\caption{(Color online) 
Upper:
The chiral condensates $\sigma$, Polyakov loops $(\ell,\bar{\ell})$,
in NLO Haar measure MFA as a function of $\mu$
at $(\beta,m_0) = (4.0,0.0)$
for the selected temperatures $T = 0.15,\ 0.2,\ 0.25$ in lattice units.
Lower:
The $\sigma$, $(\ell,\bar{\ell})$, and quark number density $\rho_q$
in NLO Haar measure MFA as a function of $\mu$
at $(\beta,m_0,T) = (4.0,0.0,0.25)$ in lattice units.
}\label{Fig:OrdEq_H_40_00_finite_mu}
\end{figure}
In the upper panel of Fig.~\ref{Fig:OrdEq_H_40_00_finite_mu},
we show the chiral condensates $\sigma/N_c$
as a function of chemical potential $\mu$
for three fixed temperatures $T = 0.15,\ 0.20,\ 0.25$.
At $T = 0.25$ (red-solid line), we find the second-order phase transition.
At lower $T \sim 0.20$ (blue-dashed line),
the chiral symmetry is partially restored with
the first-order phase transition as $\mu$ increases,
and gets completely restored
with the second-order phase transition at larger $\mu$.
As shown in the previous study \cite{NLO},
the partial chiral restoration (PCR)
emerges due to the self-consistent evaluation
of the finite $\beta$ effects for the chemical potential:
The {\em effective} chemical potential appears
as an implicit function of $\sigma$,
$\mu\to \tilde{\mu}(\sigma,\beta) = \mu - \delta\mu(\sigma,\beta)$
(see, Table \ref{Tab:renorm}),
which allows a stable equilibrium
satisfying $\sigma \sim \tilde{\mu}(\sigma)$,
leading to the PCR.
Our finding in the present study is that
the PCR is not spoiled by the Polyakov loop effects, but still exists.
As $T$ decreases, the PCR disappears and the first-order chiral transition dominates
as indicated by the $T = 0.15$ case (dashed-dotted black line).

In the lower panel of Fig.~\ref{Fig:OrdEq_H_40_00_finite_mu},
we pick up the $T = 0.25$ case from the upper panel
and show the $\mu$ dependence of $\sigma/N_c$ in a wider range.
The Polyakov loops ($\ell,\bar{\ell}$)
and the quark number density ($\rho_q/N_c$ defined by Eq.~(\ref{Eq:dens}))
are also displayed.
The Polyakov loops increase
in the chiral broken phase $\mu < \mu_c\simeq 0.59$,
and the increasing rate stays quite small
compared with the finite $T$ transition case.
In contrast, the quark number density rapidly increases
in the vicinity of the chiral phase transition.
After the transition ($\mu\geq \mu_c$),
we observe a high density system ($\rho_q \sim N_c$)
with a little quark excitation ($\ell \ll 1$).
This property as well as the possibility of two sequential transitions
associated with the PCR would be
reminiscent of the original idea of the quarkyonic phase \cite{McLerran:2007qj}.

\begin{figure}[ht]
\includegraphics[width=8.0cm]{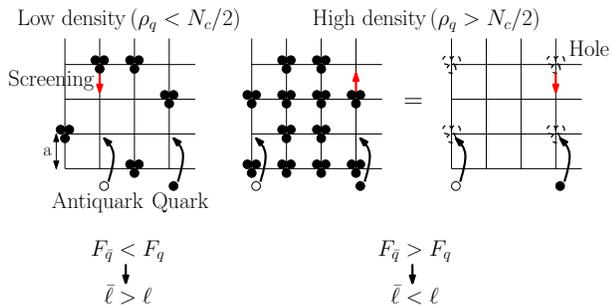}
\caption{(Color online) The half-filled and saturation.
}\label{Fig:HF-Sat}
\end{figure}
In the symmetric phase,
the Polyakov loops ($\ell,\bar{\ell}$)
start decreasing with the relation $\bar{\ell} < \ell$ as $\mu$ increases.
This would be a saturation artifact on the lattice:
As we explained above, the chiral symmetry restoration
leads to a high density system $\rho_q > N_c/2$
so that more than half of the lattice sites are filled by quarks.
Then the holes - sites without quarks - behave like antiquarks,
and the system with the quark number density $\rho_q > N_c/2$
would be identical to the system with
the antiquark number density $\rho_{\bar{q}} = (N_c - \rho_{q}) < N_c/2$.
Therefore, the excitation property of quarks and antiquarks becomes opposite
($F_q < F_{\bar{q}}$) as illustrated in Fig~\ref{Fig:HF-Sat},
and thus $\bar{\ell} < \ell$ holds.
As $\mu$ becomes larger after the half-filling,
the number of holes decreases and the degrees of freedom get frozen.
Hence the excitations of both quarks and antiquarks
are suppressed at larger $\mu$,
which results in the decreasing trend of ($\ell,\bar{\ell}$)
as functions of $\mu$.

\begin{figure}[ht]
\includegraphics[width=8.0cm]{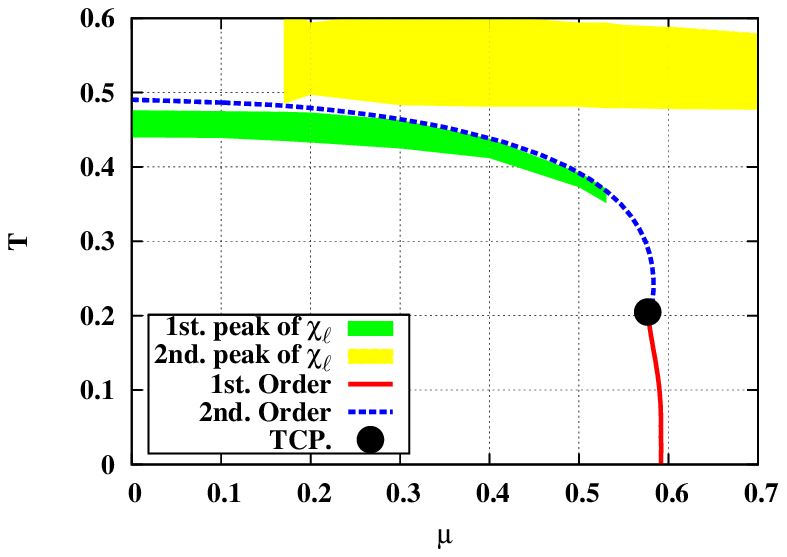}
\includegraphics[width=8.0cm]{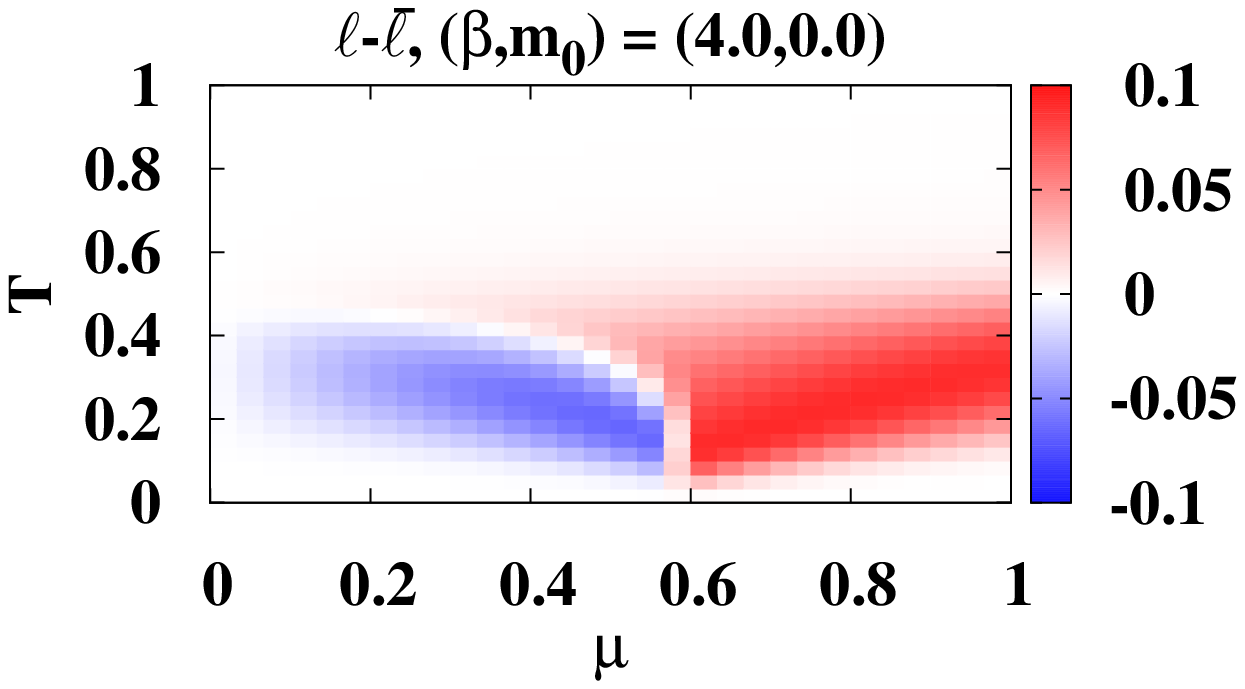}
\caption{(Color online)
Upper: The phase diagram of NLO Haar measure MFA
at $(\beta,m_0) = (4.0,0.0)$ in lattice units.
See texts for details.
Lower: The difference of the Polyakov loop and anti-Polyakov loop
in the phase diagram of the NLO Haar measure MFA.
}\label{Fig:PD_H_40_00}
\end{figure}
We show the phase diagram of NLO Haar measure MFA
in the upper panel of Fig.~\ref{Fig:PD_H_40_00}
with $(\beta,m_0) = (4.0,0.0)$.
The first-order chiral phase boundary
(red-solid line) emerges in the low $T$ region
and ends up with the tricritical point (TCP, filled black circle)
at $(\mu_{\sub{TCP}},T_{\sub{TCP}}) \simeq (0.577, 0.205)$,
from which the second-order chiral phase boundary (blue-dashed line)
sets in with increasing temperature
(The PCR emerges just below the TCP,
and invisible in the resolution of Fig.~\ref{Fig:PD_H_40_00}.
The PCR becomes visible at larger $\beta$ as seen in Fig.~\ref{Fig:PD_evol_00}).
The lower-green (upper-yellow) band
corresponds to the width of the Polyakov loop susceptibility
$\chi_{\ell}$ at 90\% of its first (second) peak height.
The first peak band depends on $\mu$ similarly
to the chiral phase boundary:
As mentioned above,
the peak seems to be associated with the chiral phase transition.
The peak strength becomes weaker with increasing $\mu$,
and disappears at $\mu\simeq 0.53$ before reaching TCP.
The second peak is almost independent of $\mu$,
and starts appearing in $\mu\gtrsim 0.17$ separately from the first peak.

The phase diagram in the Haar measure MFA
is similar to that in PQM~\cite{Herbst:2010rf}:
When the $\mu$ dependence is absent in the Polyakov loop potential in PQM,
the derivative of the Polyakov mean-field in terms of $T$ at finite $\mu$
has double peaks, which is analogous to our result shown
in the lower panel of Fig.~\ref{Fig:H_40_00_T_04}
as well as in our previous study \cite{Miura:2011kq}.
We will revisit this subject in Weiss MFA case in the next subsection.

The lower panel of Fig.~\ref{Fig:PD_H_40_00} shows the difference of
the Polyakov loop and anti-Polyakov loop
($\ell - \bar{\ell}$) in the $T-\mu$ plane.
The relation $\ell < \bar{\ell}$ 
holds in the whole $T,\mu> 0$ region in the chiral broken phase
as shown by the blue color.
The saturation effect $\ell > \bar{\ell}$ is observed
as a general tendency at large $\mu$ region in the chiral restored phase
as indicated by the red color.

As shown in Eq.~(\ref{Eq:gg_ploop_action}),
the plaquette-driven Polyakov loop action includes the $\mathcal{O}(1/g^{2/T})$ correction.
At the chiral phase boundary, this effect gives $\mathcal{O}(1/g^{2/T_c})\lesssim \mathcal{O}(1/g^{4})$.
For the consistency of the strong coupling expansion,
we have to take account of the NNLO $1/g^4$ effects for the quark sector,
which will be discussed in the later subsection.

\subsection{Weiss MFA at NLO}\label{subsec:NLO_Weiss}
We investigate the phase diagram of NLO Weiss MFA,
where the Polyakov loop fluctuations
from the mean fields $(\ell,\bar{\ell})$ are considered,
while the NNLO effects $\mathcal{O}(1/g^4)$ 
in the coupling coefficients shown in Table \ref{Tab:coupling} are ignored.
We compare the Weiss MFA results with the Haar measure MFA
to clarify the effects of the Polyakov loop fluctuations to the phase diagram.
We choose the same parameter set
as the Haar measure MFA case, $(\beta, m_0) = (4.0, 0.0)$.

\begin{figure}[ht]
\includegraphics[width=8.0cm]{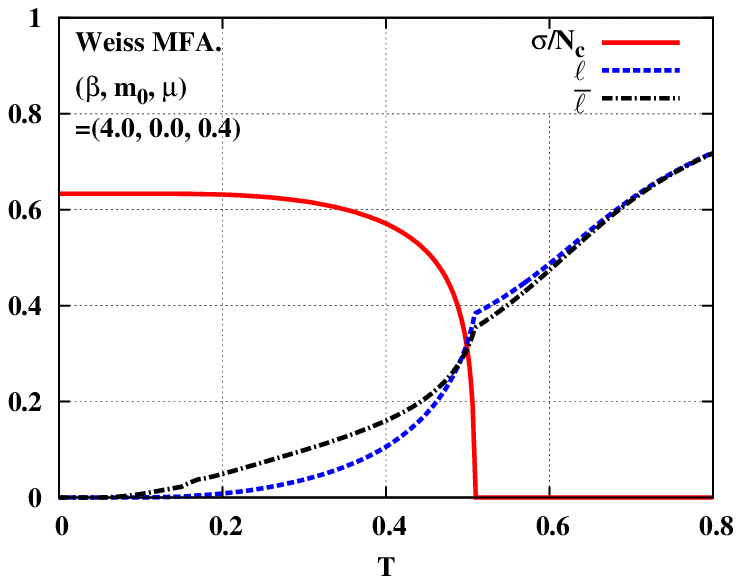}
\includegraphics[width=8.0cm]{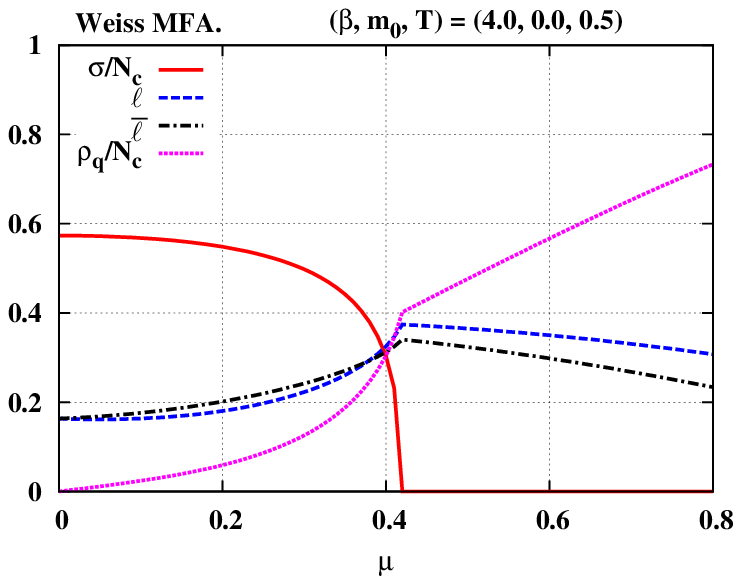}
\caption{(Color online) 
Upper:
The chiral condensates $\sigma$, Polyakov loops $(\ell,\bar{\ell})$,
in NLO Weiss MFA as a function of $T$
at $(\beta,m_0,\mu) = (4.0,0.0,0.4)$ in lattice units.
Lower:
The chiral condensates $\sigma$, Polyakov loops $(\ell,\bar{\ell})$,
and quark number density $\rho_q/N_c$
in NLO Weiss MFA as a function of $\mu$
at $(\beta,m_0,T) = (4.0,0.0,0.5)$ in lattice units.
}\label{Fig:OrdEq_W_40_00}
\end{figure}

As shown in Fig.~\ref{Fig:OrdEq_W_40_00},
$T$ or $\mu$ dependence of $(\sigma,\ell,\bar{\ell},\rho_q)$
is qualitatively the same as the Haar measure MFA results.
In the following, we concentrate on the results
which are characteristic of the Weiss MFA.

\begin{figure}[ht]
\includegraphics[width=8.0cm]{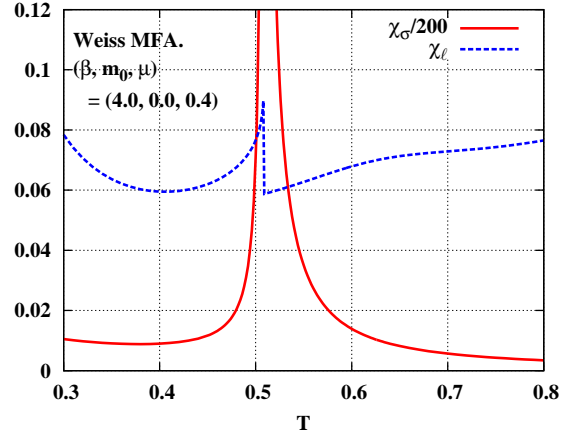}
\caption{(Color online) 
The chiral and Polyakov loop susceptibilities
($\chi_{\sigma},\chi_{\ell}$)
in NLO Weiss MFA as a function of $T$
at $(\beta,m_0,\mu) = (4.0,0.0,0.4)$ in lattice units.
For a comparison, the chiral susceptibility $\chi_{\sigma}$
has been multiplied by 1/200.
}\label{Fig:Suscept_W_40_00_T_04}
\end{figure}

In Fig. \ref{Fig:Suscept_W_40_00_T_04},
we show the chiral and Polyakov loop susceptibilities
($\chi_{\sigma},\chi_{\ell}$)
at finite chemical potential $\mu = 0.4$
as a function of temperature $T$.
Two peaks are almost degenerated,
and the width of $\chi_{\ell}$ is sharper than the Haar measure MFA case.
We do not see the second ($Z_3$ associated) peak in the chiral symmetric phase
in sharp contrast to the Haar measure MFA case.

\begin{figure}[ht]
\includegraphics[width=8.0cm]{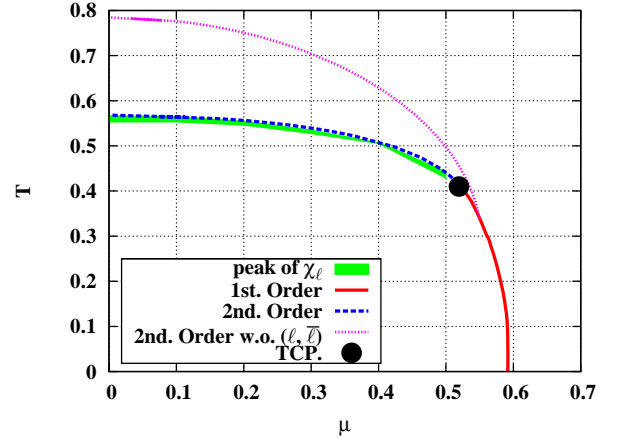}
\caption{(Color online) The phase diagram
at $(\beta,m_0) = (4.0,0.0)$ in NLO with Weiss MFA in lattice units.
See texts for details.
}\label{Fig:PD_W_40_00}
\end{figure}

In Fig.~\ref{Fig:PD_W_40_00},
we show the phase diagram of NLO Weiss MFA
with $(\beta,m_0) = (4.0,0.0)$.
We find two qualitative differences
between the NLO Weiss MFA and NLO Haar measure MFA results:
First, the peak of $\chi_{\ell}$
(green-band showing the width of $\chi_{\ell}$ at 90\% of the peak height)
is more strongly locked to the chiral phase boundary in Weiss MFA
than the Haar measure MFA case.
Second, the remnant of the $Z_3$ dynamics such as the yellow band in
Fig.~\ref{Fig:PD_H_40_00} does not appear at any $\mu$ in the Weiss MFA case.
As explained after Eq.~(\ref{Eq:FPol_Weiss}) in the previous section,
the plaquette-driven Polyakov loops are combined into the quark determinant
and coupled to the dynamical quark effects via the $U_0$ path integral.
Then, the Weiss MFA does not admit the remnant of the $Z_3$ symmetry
in sharp contrast to the Haar measure MFA and
many other chiral effective models~\cite{Fukushima:2003fw,Fukushima:2008wg,Herbst:2010rf}.
It is sometimes argued that
the chiral and deconfinement dynamics might be separated at finite $\mu$~\cite{Fukushima:2013rx},
but the Weiss MFA does not support the isolated deconfinement dynamics from the chiral phase boundary.

Here, we comment on the recent phase diagram study by
the PQM model \cite{Herbst:2010rf}.
In this model,
a $\mu$ dependence was assumed in the Polyakov loop effective potential
based on the phenomenological insights
to describe the backreaction of the quark-matter to the Polyakov loops at finite density.
This prescription
led to a stronger locking between the peak of $d\ell/dT$
and the chiral crossover line,
and the double peak structure of $d\ell/dT$ disappeared.
These phenomena would be analogous to our findings in the Weiss MFA.
We stress that the Weiss MFA effective potential
directly results from the path integral in the lattice QCD
without additional assumptions.
This would be the advantage of the SC-LQCD based effective potential.

We shall consider the formal limit $(\ell,\bar{\ell})\to 0$
in the effective potential of Weiss MFA $\mathcal{F}_{\mathrm{eff}}^{\sub{W}}$:
In the second line of Eq.~(\ref{Eq:FDetW}),
the thermal excitations (see Table \ref{Tab:FDetW})
carrying a quark number $0$ ($I=\mathrm{MMM},\mathrm{MQ\bar{Q}}$)
and $\pm 3$ ($I=\mathrm{B},\mathrm{\bar{B}}$) remains 
and the $\mathcal{F}_{\mathrm{eff}}^{\sub{W}}$ reduces into
the effective potential
which has been derived in our previous study \cite{NLO}.
We express the reduced effective potential as $\mathcal{F}_{\mathrm{eff}}^{\sub{NLO}}$,
and the results obtained by using $\mathcal{F}_{\mathrm{eff}}^{\sub{NLO}}$
will be referred to as {\em NLO without Polyakov loops} in the later discussions.
See Eq.~(\ref{Eq:app_NLO}) for the expression of $\mathcal{F}_{\mathrm{eff}}^{\sub{NLO}}$.
Needless to say, the $\mathcal{F}_{\mathrm{eff}}^{\sub{NLO}}$
does not implement the Polyakov loop dynamics.
By comparing the Weiss NLO MFA and the NLO without Polyakov loops,
the Polyakov loop effects become more transparent.

In Fig.~\ref{Fig:PD_W_40_00},
we compare the chiral phase boundary of the NLO Weiss MFA
and the NLO without Polyakov loops.
The second-order phase boundary of the NLO Weiss MFA (blue-dashed line)
is found in lower $T$ region than that of
NLO without Polyakov loop (magenta-dotted line).
As $\mu$ becomes larger,
two phase boundaries get closer to each other
and degenerate in the vicinity of the TCP.
The first-order phase boundary
is almost independent of the Polyakov loop effects.
This is understood as follows.
As explained in the previous section,
the plaquette-driven Polyakov loops gives
the contribution of $\mathcal{O}([1/g^2]^{1/T_c(\mu)})$.
At larger $\mu$, this factor decreases because the $T_c(\mu)$ does,
and thus the Polyakov loop effects becomes higher order effects
of the strong-coupling expansion, and thereby suppressed.

Compared with the Haar measure MFA,
the transition temperature $T_c(\mu)$ in the Weiss MFA becomes somewhat larger.
Then, the effect of the plaquette-driven Polyakov loops for $\beta = 4.0$ is maximally
$\mathcal{O}([1/g^2]^{1/T_c(\mu = 0)}) = \mathcal{O}(1/g^{3.3})$,
which is larger than the NNLO effects $1/g^4$.
Thus, the present NLO approximation for the quark sector
is consistent with respect to the order counting of the strong coupling expansion,
at least for $\beta \lesssim 4.0$.

\begin{figure}[!ht]
\includegraphics[width=8.0cm]{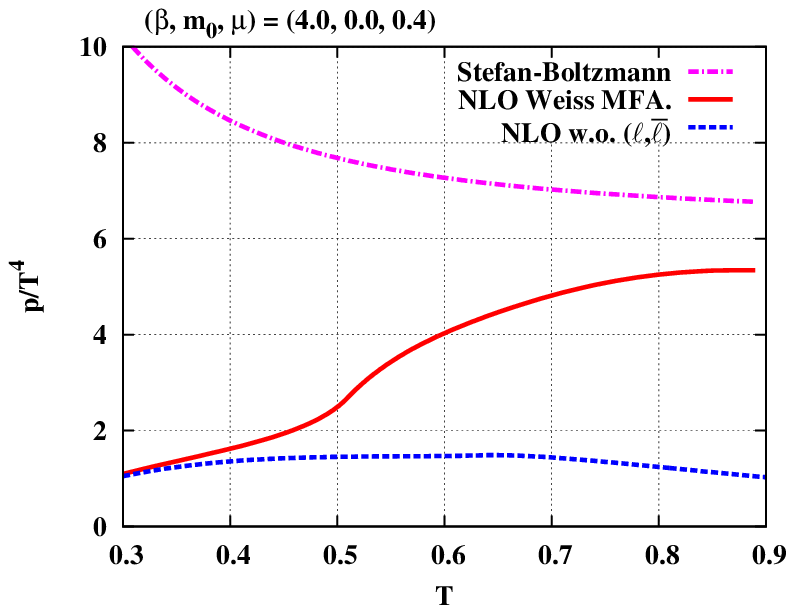}
\includegraphics[width=8.0cm]{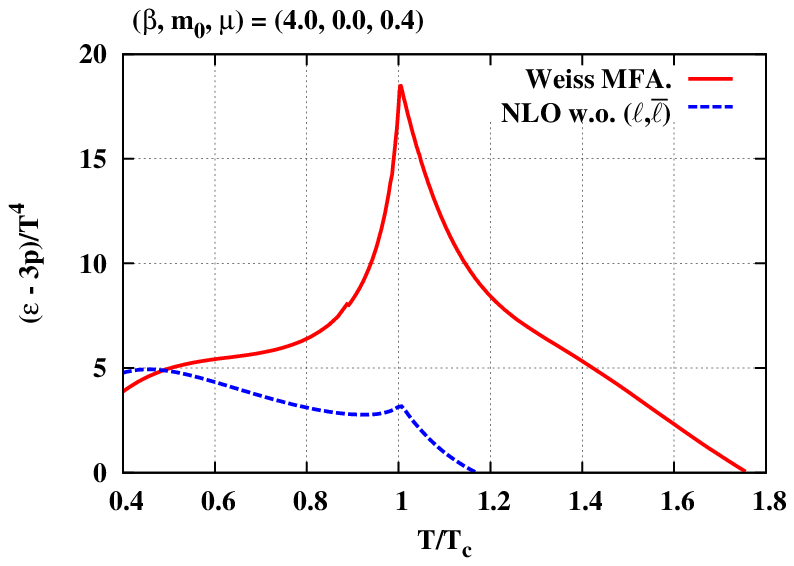}
\includegraphics[width=8.0cm]{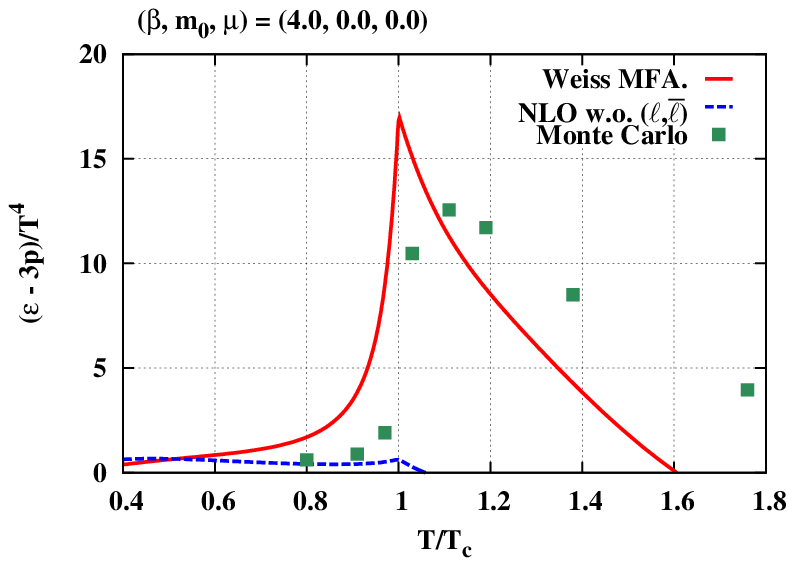}
\caption{(Color online) 
The pressure (upper) and interaction measure (middle and lower)
normalized by $T^4$ 
in ``NLO Weiss MFA'' and ``NLO without Polyakov loops''
as a function of $T$ 
at $(\beta,m_0,\mu) = (4.0,0.0,0.4)$ (upper and middle),
and $(\beta,m_0,\mu) = (4.0,0.0,0.0)$ (lower).
In the upper panel, the horizontal axis $T$ is in lattice units.
In the middle and lower panels, the horizontal axis is normalized
by critical temperature $T_c$.
In the lower panel, we have quoted the Monte Carlo results
\protect\cite{Engels:1996ag}.
}\label{Fig:TH_W_40_00_T_04}
\end{figure}
Next, we investigate the thermodynamic quantities in the Weiss MFA.
In the upper panel of Fig.~\ref{Fig:TH_W_40_00_T_04},
we show the normalized pressure $p/T^4$ as a function of $T$
at chemical potential $\mu = 0.4$,
the same condition as Fig.~\ref{Fig:Suscept_W_40_00_T_04}.
In NLO Weiss MFA,
the $p/T^4$ (red-solid line) becomes significantly larger
at $T\gtrsim T_c\simeq 0.507$ 
and closer to the Stefan-Boltzmann result
\begin{align}
\frac{p_{\mathrm{sb}}}{T^4} =
\frac{N_fN_c}{6}
\biggl[\frac{7\pi^2}{30}
+ \frac{\mu^2}{T^2}
+ \frac{1}{2\pi^2}\frac{\mu^4}{T^4}
\biggl]
+ \frac{(N_c^2 - 1)\pi^2}{45}\ .
\end{align}
We do not see such a large enhance of $p/T^4$ in the case of
NLO without Polyakov loops (blue-dashed line).
Thus, the Polyakov loop plays an essential role
to realize the pressure enhancement which is expected in
the QGP phase at high $T$.
More specifically,
the pressure enhancement is attributed to the increase
of Polyakov loop thermal excitations
$\mathcal{P}_n^I(\sqrt{\ell\bar{\ell}})$ (see Table \ref{Tab:FDetW})
included in the Weiss MFA effective potential
(\ref{Eq:FDetW})-(\ref{Eq:Pn}).
This result should be compared with that in the PQM model,
where the pressure is rather suppressed by Polyakov loops~\cite{Herbst:2010rf}.
The different role of Polyakov loops is understood as follows.
First, we recall that a usual NJL (QM) does not implement a confinement dynamics
since quarks are introduced without gauge interactions.
When Polyakov loop effects are introduced, giving PNJL (PQM),
the Boltzmann factors for quark thermal excitations
in the effective potential~\cite{Fukushima:2008wg,Herbst:2010rf}
are multiplied by the Polyakov loop mean-field $\ell$,
which acts as a suppression factor of the quark thermal excitations at low $T$.
In this sense, the role of the Polyakov loop is to {\em confine} quarks at low $T$ in PNJL and PQM,
and therefore, suppresses the pressure.
By comparison in SC-LQCD, the link integrals admit only color-singlet hadronic states contributing
to the effective potential. As a result, the thermal excitations carrying the quark number
$N_{\mathrm{Q}}^{I} = \pm 1$ (quark and antiquark excitations) and $\pm 2$ (diquark and anti-diquark excitations)
in Table~\ref{Tab:FDetW} can emerge only when the Polyakov loop mean-fields are taken account.
In this sense, the role of the Polyakov loop is to {\em deconfine} quarks at high $T$ in SC-LQCD, and enhances the pressure.
Thus, Polyakov loops play different roles in the SC-LQCD and PNLO (PQM).

In the middle panel of Fig.~\ref{Fig:TH_W_40_00_T_04},
we show the interaction measure $\Delta = (\epsilon - 3p) / T^4$ as a function of $T$
at chemical potential $\mu = 0.4$.
In NLO Weiss MFA,
the $\Delta$ has a large peak
in the vicinity of the chiral phase transition $T\sim T_c$
as expected with regards to the increasing scale asymmetry
in the strongly interacting quark-gluon plasma (sQGP).
This should be compared with
the result obtained in NLO without Polyakov loops (dashed-blue line)
staying small and showing just a tiny bump structure at $T\sim T_c$.
In the lower panel of Fig.~\ref{Fig:TH_W_40_00_T_04},
we compare our results on the interaction measure
at vanishing of chemical potential
with those obtained in the Monte Carlo simulations
(four flavor, the chiral limit is taken) \cite{Engels:1996ag}.
The Monte Carlo results (green boxes) show the drastic increase
in the vicinity of the chiral phase transition.
This feature is qualitatively reproduced by
the NLO Weiss result (red-solid line),
but not in the NLO without Polyakov loops (blue-dashed line).
Around $T = T_c$, a singular behavior in the derivative of $(\epsilon - 3p) / T^4$
with respect to $T$ is seen only in the chiral limit.
This behavior is associated with the second-order chiral phase transition
as mentioned in Sec. \ref{subsec:NLO_Haar}.

\subsection{Quark mass dependence}\label{subsec:qmass}
In the previous subsections,
we have studied the phase diagram in the chiral limit $m_0 = 0$.
In this subsection,
we investigate the $m_0$ dependence of
the chiral and Polyakov loop susceptibilities.
We choose the same parameter set of $\beta = 4.0$ and $\mu = 0.4$
as previous subsections.

\begin{figure}[ht]
\includegraphics[width=8.0cm]{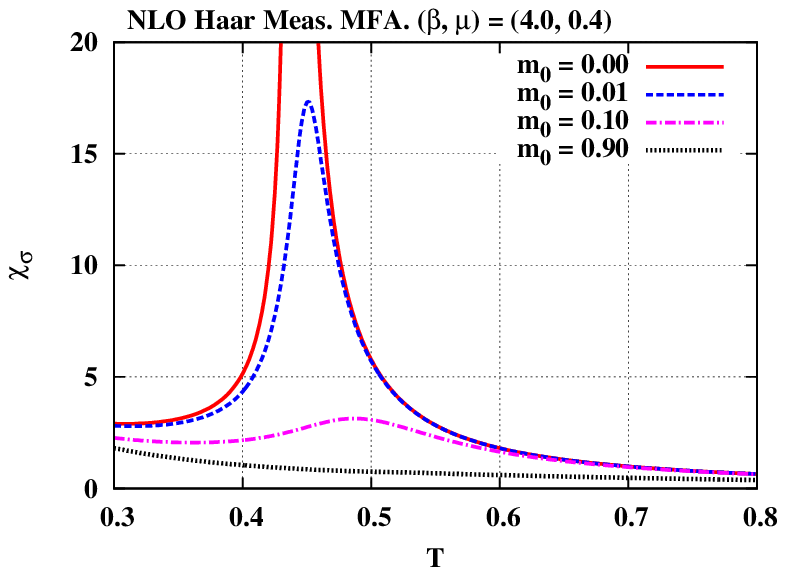}
\includegraphics[width=8.0cm]{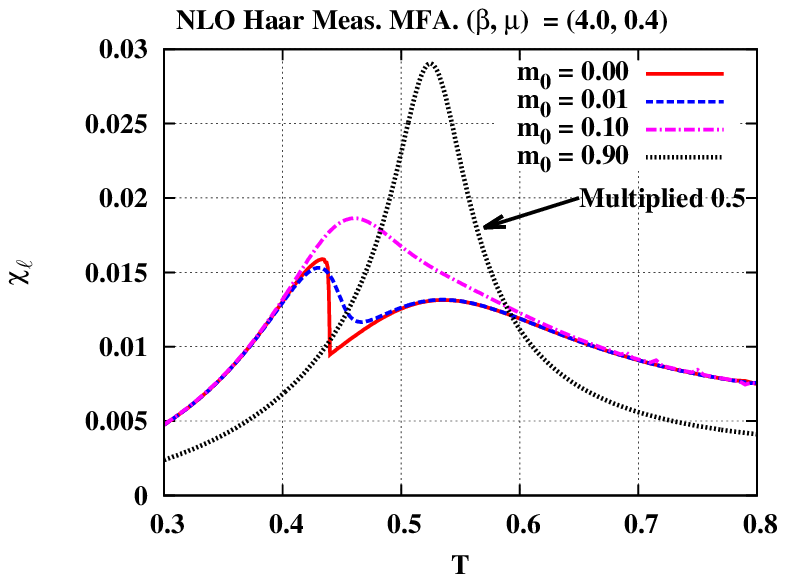}
\caption{(Color online)
The chiral (upper) and Polyakov loop (lower) susceptibilities
as a function of $T$ for various bare quark mass $m_0$
at $(\beta,\mu) = (4.0,0.4)$  in NLO Haar measure MFA.
All quantities are in lattice units.
}\label{Fig:Suscept_HT_40_xx_T_04}
\end{figure}
\begin{figure}[ht]
\includegraphics[width=8.0cm]{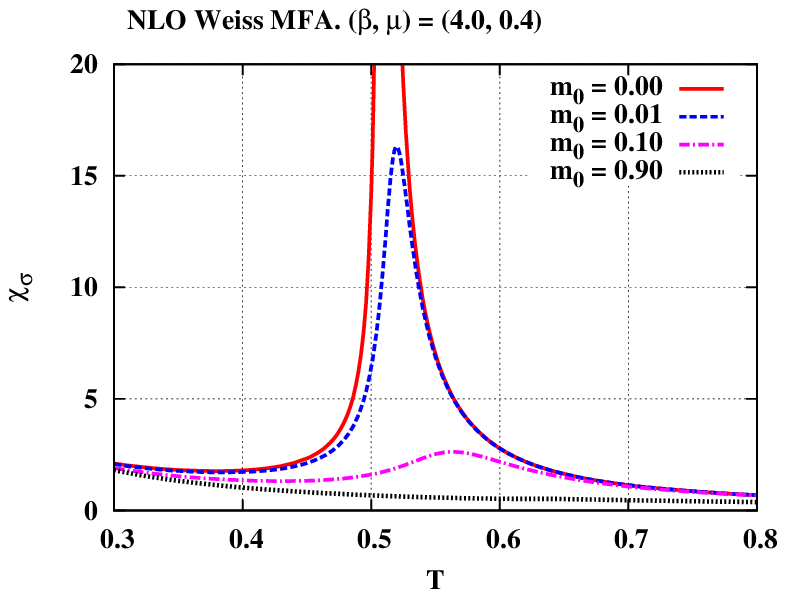}
\includegraphics[width=8.0cm]{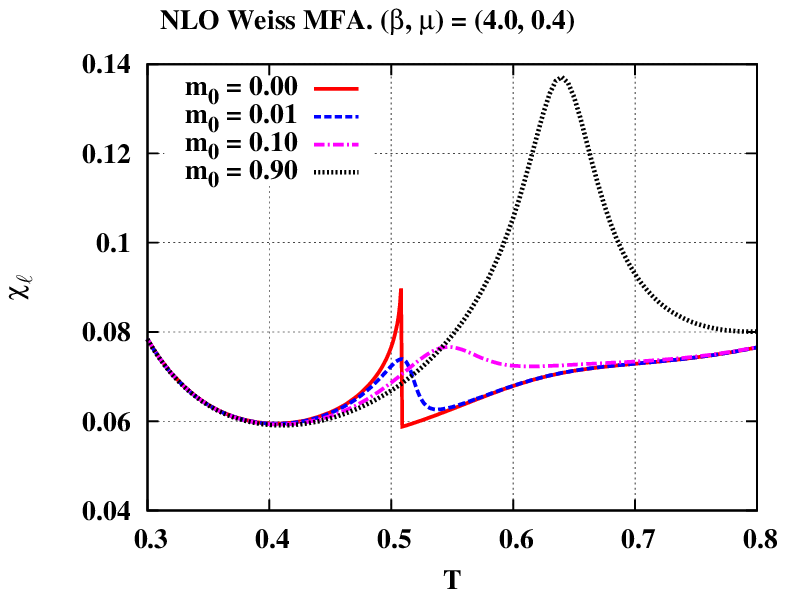}
\caption{(Color online)
The chiral (upper) and Polyakov loop (lower) susceptibilities
as a function of $T$ for various bare quark mass $m_0$
at $(\beta,\mu) = (4.0,0.4)$  in NLO Weiss MFA.
All quantities are in lattice units.
}\label{Fig:Suscept_WT_40_xx_T_04}
\end{figure}

In the upper panel of Fig.~\ref{Fig:Suscept_HT_40_xx_T_04},
we show the chiral susceptibility $\chi_{\sigma}$
of the NLO Haar measure MFA
as a function of $T$ for various bare quark mass $m_0$
at $(\beta,\mu) = (4.0,0.4)$.
The peak position defines the chiral crossover temperature at finite $m_0$.
The chiral dynamics becomes weaker
as indicated by the attenuating peak with increasing $m_0$.
In the lower panel of Fig.~\ref{Fig:Suscept_HT_40_xx_T_04},
we show the Polyakov loop susceptibility $\chi_{\ell}$
in the same condition as the upper panel.
The double peak structure
which we have shown in the chiral limit in Subsec~\ref{subsec:NLO_Haar}
evolves into a single peak with increasing $m_0$.
The single peak grows up in the heavy mass region $m_0 = 0.9$,
and comes to be responsible for the $Z_3$ crossover.
Consistently, the chiral susceptibility does not show any signal there
as shown in the upper panel.

We notice that the $Z_3$ peak of $\chi_{\ell}$ at $m_0 = 0.9$ 
locates at the almost same temperature as
the second peak appearing in the small mass region $m_0 \lesssim 0.01$.
This implies that the second peak originates from the remnant of the $Z_3$ dynamics.
In fact, the approximate $Z_3$ symmetry remains even in the chiral limit
in the effective potential of the Haar measure MFA:
The $Z_3$ symmetric (Haar measure) term
$\mathcal{R}_{\mathrm{Haar}}$ in Eq.~(\ref{Eq:R_Haar})
has a large contribution and
does not couple to the dynamical fermion effects
$\mathcal{R}_{q}$ in Eq.~(\ref{Eq:FDet_Haar_app}),
so that the former effect is not horribly spoiled by the latter.
The result is consistent with our previous work~\cite{Miura:2011kq}.

This should be compared with NLO Weiss MFA results,
Fig.~\ref{Fig:Suscept_WT_40_xx_T_04}.
The chiral susceptibility $\chi_{\sigma}$ (upper panel) is qualitatively
the same as the Haar measure result,
while the Polyakov loop susceptibility
$\chi_{\ell}$ (lower panel) differs:
The Weiss MFA does not lead to
the double-peak structure in $\chi_{\ell}$ for any $m_0$.
Thus, the scenario with the double-peak, or equivalently, the deconfinement separated from the chiral phase boundary
at high density would be less supported within the present approximation.
To extract a definite conclusion on
the relation between two susceptibilities $\chi_{\sigma,\ell}$,
we need to investigate the higher-order effects on the Polyakov loops.

It is worth mentioning that the Polyakov loop effective potential
in the Haar measure MFA, Eq.~(\ref{Eq:FPol_Haar})
is similar to one of the popular choices of the potential in the PNJL model~\cite{Fukushima:2003fw,Fukushima:2008wg}
or PQM models. They could in principle contain the remnant of $Z_3$ dynamics as the Haar measure MFA does.
As explained in the previous subsection,
the recent work based on PQM assumed
a certain $\mu$ dependence to the coefficients in the Polyakov loop effective potential~\cite{Herbst:2010rf}.
This gives a phenomenological implementation of a back reaction from dynamical quark effects.
The Weiss MFA effective potential (especially Eq.~(\ref{Eq:FDetW}))
proposes the lattice QCD based solution for the quark back reaction to the Polyakov loops,
and opens a possibility to upgrade the PNJL and PQM models
so that they account for the Polyakov loop and quark degrees of freedom more systematically.
To invent such a model based on the Weiss MFA should be one of the future works.

\subsection{
Phase diagram evolution with increasing $\beta$}
\label{subsec:PD_evol}
So far, we have studied the phase diagram at a fixed coupling, $\beta=4.0$.
In this subsection, we investigate the phase diagram 
for various lattice coupling ranging $0.0\leq\beta\leq 6.0$,
while we keep the vanishing bare quark mass $m_0 = 0$.
For the chiral phase transition temperature
at vanishing chemical potential $T_{c,\mu = 0}$,
the lattice MC data with one species of staggered fermion
are available \cite{deForcrand:2009dh,Forcrand,MC4,Engels:1996ag}
and are compared with $T_{c,\mu = 0}$ evaluated
in the strong-coupling expansion~\cite{NLO,Nakano:2009bf,PNNLO-T}.
We extend our analyses up to $\beta = 6.0$,
for which the physical scale of $(T_c,\mu_c)$ can be extracted
by utilizing the lattice spacing result in Ref.~\cite{Jolicoeur:1983tz}.

\begin{figure}[tbhp]
\includegraphics[width=8.0cm]{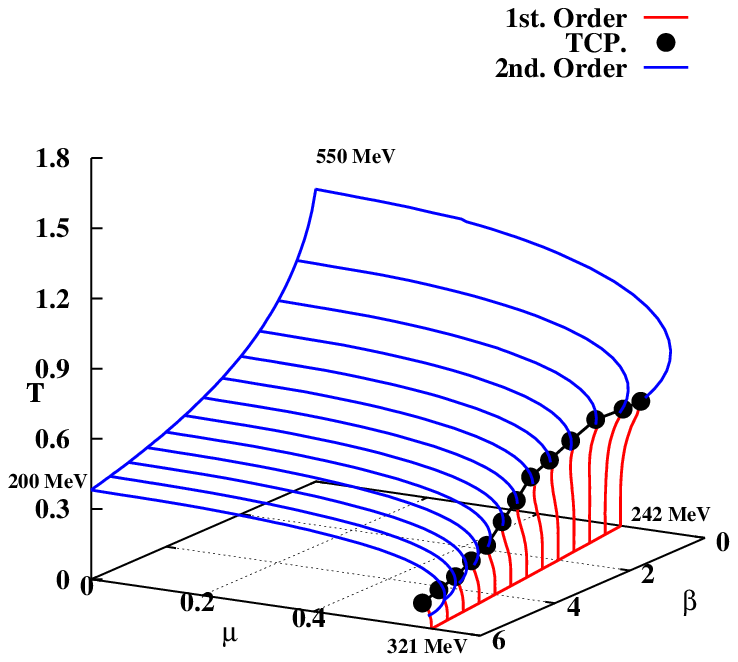}
\includegraphics[width=8.0cm]{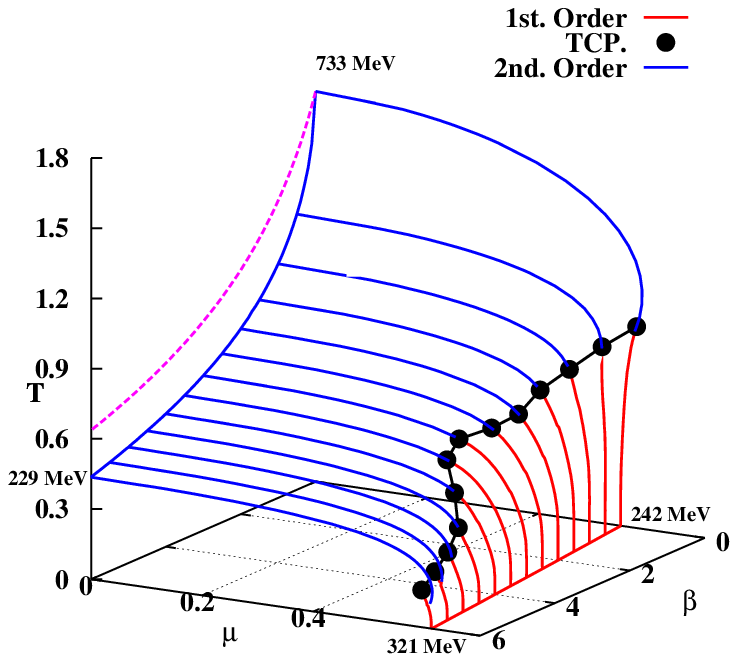}
\caption{(Color online) The phase diagram evolution
with the increasing lattice coupling $\beta$ at NLO
Haar measure (upper) and Weiss (lower) MFA
in the chiral limit.
We have quoted $a^{-1}(\beta=0)=440$ (MeV)
and $a^{-1}(\beta=6)=524$ (MeV)
from Ref.~\cite{Jolicoeur:1983tz}.
In Weiss MFA (lower),
we have also shown our previous result on
the transition temperature at $\mu=0$
without effects of plaquette-driven Polyakov loops
\protect\cite{NLO} by the magenta-dashed line.
}\label{Fig:PD_evol_00}
\end{figure}

In the upper panel of Fig.~\ref{Fig:PD_evol_00},
we show the phase diagram evolution with increasing $\beta$
in the case of NLO Haar measure MFA.
In the whole range of $0.0\leq\beta\leq 6.0$,
the chiral phase transition is a first-order
in the low temperature region,
and it evolves into the second-order at higher $T$ via TCP.
The transition temperature at $\mu=0$ ($T_{c,\mu=0}$) acquires
much larger modification with increasing $\beta$
than the transition chemical potential at $T=0$ ($\mu_{c,T=0}$).
Resultantly, the ratio $R=\mu_{c,\sub{T}=0}/T_{c,\mu=0}$
which characterize the shape of the chiral phase boundary
is greatly enhanced.
For $\beta\geq 4$, the first-order transition line
goes inside of the second-order transition line near the TCP,
and the PCR explained in the previous subsection emerges between two lines.

In the lower panel of Fig.~\ref{Fig:PD_evol_00},
we show the phase diagram evolution of the NLO Weiss MFA
in the chiral limit $m_0 = 0.0$.
The results are qualitatively same as the Haar measure MFA case.

\begin{figure}[ht]
\includegraphics[width=8.0cm]{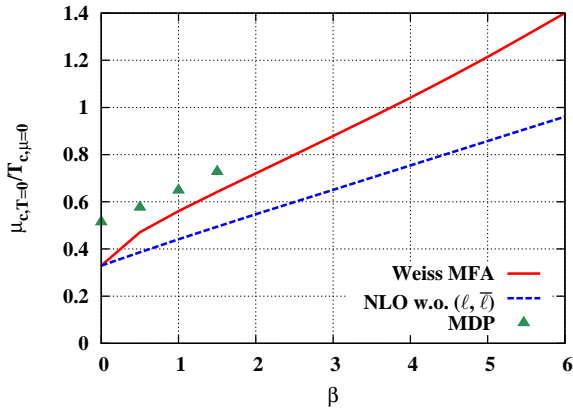}
\caption{(Color online) The ratio $R=\mu_{c,\sub{T}=0}/T_{c,\mu=0}$
as a function of the lattice bare coupling $\beta=2N_c/g^2$.
For a comparison,
we show the MDP results for $R$
in Ref.~\protect\cite{deForcrand:2014tha}:
We read off the $\mu_{c,T=0}\simeq 0.72$ from Fig. 6
in Ref.~\protect\cite{deForcrand:2014tha},
and quoted the $T_c(\beta)$ evaluated by using
the ``exponential extrapolation''
to calculate the $R$.
}\label{Fig:R_beta_00}
\end{figure}
We compare the ratio $R=\mu_{c,\sub{T}=0}/T_{c,\mu=0}$ of NLO Weiss MFA
to that obtained in the ``NLO without Polyakov loops''.
At $\beta = 4.0\ (6.0)$,
the former (red-solid line) in Fig.~\ref{Fig:R_beta_00} becomes
1.38 (1.46) times larger than the latter (blue-dashed line).
Thus the ratio $R$ becomes larger by the Polyakov loop effects.
Next, we compare our $R$
with those obtained by the Monomer-Dimer-Polymer (MDP) simulation
\cite{deForcrand:2014tha}.
The MDP (green triangles in Fig.~\ref{Fig:R_beta_00})
gives a somewhat larger $R$ than our MFA result in the strong-coupling limit,
and becomes closer to the NLO Weiss MFA at finite $\beta$.
The increasing $R$ at larger $\beta$
is a common trend in both MFA and MDP,
and preferable to be consistent with
a realistic QCD phase diagram.

In both Haar measure MFA and Weiss MFAs,
the TCP tends to go into low $T$ region with increasing $\beta$,
and the second-order chiral phase boundary becomes dominant.
However, the TCP and PCR evolution at large $\beta$ in Fig.~\ref{Fig:PD_evol_00} may be modified
by a number of effects which are missing in the present mean-field framework at NLO;
(1) fluctuation degrees of freedom from mean-fields,
(2) effects of higher-order of the strong-coupling expansion, and
(3) chiral anomaly effects.
In the following, we discuss these corrections in relation to the continuum limit.

The fluctuation effects become important in critical phenomena and
may give a non-negligible correction to the TCP and PCR obtained in MFA
even at a fixed order of the strong-coupling expansion. This is the ambiguity (1).
However, at least for the strong-coupling region $\beta \leq 1.5$,
the basic property of the TCP shown in this work would be stable against the fluctuations;
our results show that the TCP {\em exists} and it is almost independent of $\beta$,
which is shown to remain intact even including the fluctuation effects~\cite{MDP-NLO,SC-LQCD-MC}.
For the evolution of TCP/PCR at larger $\beta$, only mean-field results
(the present and previous works~\cite{NLO,Nakano:2009bf}) are available,
and it is desirable to investigate the fluctuation effects in the near future.

If we {\em assume} that the (T)CP (and thereby, the first-order chiral phase transition)
remains in the phase diagram in the continuum limit, the SC-LQCD including the fluctuation effects
may have a contact to the critical phenomena expected in the continuum limit.
This relies on the following reasoning.
The SC-LQCD with one species of the staggered fermion has $O(2)$ symmetry at finite lattice coupling,
while the massless two-flavor QCD in the continuum has $O(4)$.
Since the sign of the relevant critical exponent in $O(2)$ is the same as that in $O(4)$,
the ratio of various cumulants for the net baryon number
($\chi_{\mu}^{(n)} \propto \partial^n \log \mathcal{Z}/\partial (N_c \mu)$, $\mathcal{Z}=$partition function)
would be similar to each other.
The cumulant ratio has been investigated only in the strong-coupling limit~\cite{Ichihara:2015bcq}.
The finite coupling extension is, in principle, possible by combining the present study
with the auxiliary-field Monte Carlo formulation~\cite{SC-LQCD-MC}.

Let us move on to the ambiguity (2), effects of higher-order of the strong-coupling expansion.
We first comment on the remarkable properties at NLO;
the first-order phase boundary diminishes with increasing $\beta$ as shown in Fig.~\ref{Fig:PD_evol_00},
and this trend becomes rather significant at finite quark mass. For example in the Haar measure MFA,
we found that the CP associated angle $\arctan(T_{\sub{CP}}/\mu_{\sub{CP}})$
at $\beta = 4.0$ is $0.34$ at $m_0 = 0.0$ and down to $0.31$ at $m_0 = 0.05$.
Thus, both of the increasing $\beta$ and nonzero $m_0$ disfavor the first-order transition at NLO.
The question is a fate of the above properties with higher-orders.
To shed light on this, we quote the LQCD-MC results on the chiral critical surface~\cite{deForcrand:2008vr}
in the $\mu$-extended Columbia plot, where the surface evolution at finite $\mu$
implies the absence of the CP at physical point mass.
Thus, the properties at NLO explained above seems to be compatible
to the LQCD-MC results including all order of $\beta$. This implies that the qualitative feature of
the (T)CP at NLO would not be horribly changed by higher order effects.
Of course, this naive expectation should be confirmed by investigating the higher-orders in future works.
We note that the absence of CP at physical point mass does not necessarily means
the absence of critical phenomena, and the above discussion of the cumulant ratio
for the ambiguity (1) can be compatible to the discussion here.

According to the effective model~\cite{Pisarski:1983ms},
the chiral phase transition in the $N_f = 4 > 2$ system
is predicted to be the first-order due to the chiral anomaly in the chiral limit.
In the SC-LQCD with staggered fermions, however, the chiral anomaly is cancelled out
among the species doublers and therefore missing in the present study.
This is the ambiguity (3), and the anomaly effect may modify the properties
of the TCP/PCR presented in this work. To shed light on this issue, we need to develop
the SC-LQCD formulation with overlap fermions.
We find some pioneering works~\cite{SCLQCD-OLF};
it was argued that a massive flavor-singlet pseudoscaler meson could appear in SC-LQCD
from a Jacobin term associated with a chirally-covariant transformation
of the path-integral measure over quark fields. This was interpreted as a solution to the U(1) problem
in the SC-LQCD context~\cite{SCLQCD-OLF}. Thus, the Jacobian term seems to play an essential role to remedy
the anomaly problem in SC-LQCD but has not been investigated
in the literature of finite $T$ and/or $\mu$ ({\em c.f.}~\cite{Yu:2005eu}).
This should also be the subject studied in future.

Finally, we estimate the $(T_c,\mu_c)$ in physical units
by quoting the lattice spacing scale $a^{-1}(\beta=0)=440$ (MeV)
and $a^{-1}(\beta=6)=524$ (MeV)
from the zero temperature strong-coupling expansion~\cite{Jolicoeur:1983tz}.
In Haar measure MFA,
we find $(T_{c,\mu=0},\mu_{c,T=0})\simeq (550,242)$ (MeV)
in the strong-coupling limit,
and $(T_{c,\mu=0},\mu_{c,T=0})\simeq (200,321)$ (MeV)
at $\beta=6$. 
In Weiss MFA, 
we find $(T_{c,\mu=0},\mu_{c,T=0})\simeq (733,242)$ (MeV)
in the strong-coupling limit,
and $(T_{c,\mu=0},\mu_{c,T=0})\simeq (229,321)$ (MeV)
at $\beta=6.0$.
Although the flavor-chiral structure of the present system
differs from the real-life QCD,
it is still interesting that
the transition temperature of SC-LQCD gets closer
to the realistic one $T_c^{\sub{MC}} = 145-195~\mathrm{MeV}$ \cite{Borsanyi:2010bp}.

\medskip

\subsection{Haar measure MFA at NNLO}\label{subsec:NNLO_Haar}
We investigate the phase diagram in the NNLO Haar measure MFA,
where the $\mathcal{O}(1/g^4)$ terms
in the coupling coefficients (Table \ref{Tab:coupling}) are considered.
We adopt the same parameter set $(\beta, m_0) = (4.0, 0.05)$
as that adopted in the previous work~\cite{PNNLO-T}.
We investigate the property of the chiral condensates and the Polyakov loops
at intermediate and high density region: $\mu = 0.4$ and $0.7$.
We compare the NNLO phase diagram with the NLO one,
and studies the impact of the NNLO corrections.

\begin{figure}[htb]
\includegraphics[width=8.0cm]{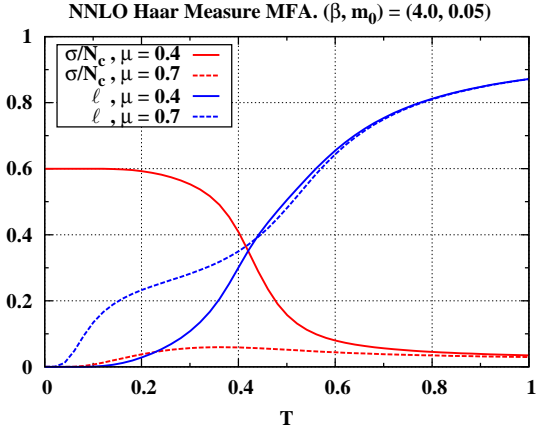}
\caption{(Color online) Chiral condensates and Polyakov loops in the NNLO Haar measure MFA
for $\beta = 4.0$ with $m_0 = 0.05$ as a function of temperature at the two chemical potentials $\mu=0.4$ and $0.7$.
All quantities are in lattice units.}
\label{Fig:Tdep-NNLO}
\end{figure}
In Fig.~\ref{Fig:Tdep-NNLO}, we show the chiral condensate
and Polyakov loop as a function of $T$ for the lattice coupling $\beta = 4.0$.
First, we consider the $\mu = 0.4$ cases.
At low $T$, the chiral symmetry is spontaneously broken ($\sigma/N_c\gg m_0 = 0.05$, red-solid line),
and the quarks are confined ($\ell \ll \mathcal{O}(1)$, blue-solid line).
At high $T$, the chiral symmetry gets restored
up to the finite bare mass effect ($\sigma/N_c\to \mathcal{O}(m_0) \sim 0.05$),
and the quarks becomes deconfined ($\ell \sim \mathcal{O}(1)$).
The chiral condensate rapidly but smoothly decreases with increasing $T$,
which indicates the chiral crossover rather than the phase transition.
At larger chemical potential $\mu = 0.7$,
the chiral condensate (red-dashed line)
is small and comparable to the bare quark mass $m_0 = 0.05$
in all $T$ region, and thus the chiral crossover is absent.

In Fig.~\ref{Fig:Tdep-NNLO}, 
we find the clear difference in the Polyakov loop $\ell$ at $\mu = 0.7$ and $0.4$;
the former (blue-dashed line) starts increasing even at a tiny (nonzero) temperature
where the latter (blue-solid line) still remains small.
This can be understood in terms of the presence/absence of the spontaneous breaking of the chiral symmetry;
at $\mu = 0.4$, the broken chiral symmetry leads to the dynamical quark mass
and suppresses the thermal excitation of the quarks, while at $0.7$, there is no suppression
due to the symmetry restoration.
Thus, the relatively large $\ell$ at low temperature can be a characteristic feature at high density phase.
At higher $T$, $\ell$ at $\mu = 0.7$ becomes comparable with that at $\mu = 0.4$.

\begin{figure}[ht]
\includegraphics[width=8.0cm]{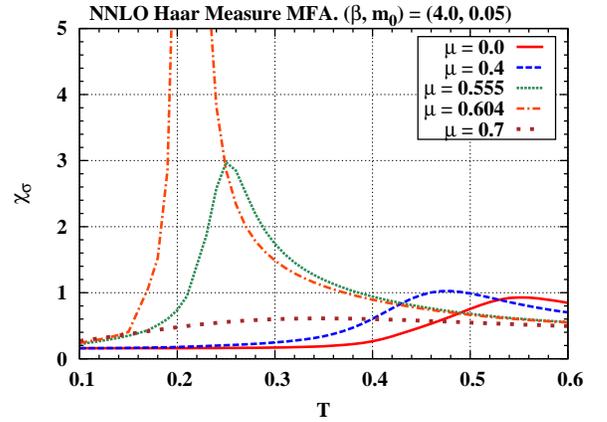}
\caption{(Color online) The chiral susceptibility in NNLO Haar measure MFA with $m_0 = 0.05$
as a function of temperature for various $\mu$.
All quantities are in lattice units at the coupling $\beta = 4.0$.}
\label{Fig:sus-NNLO}
\end{figure}
In Fig.~\ref{Fig:sus-NNLO},
we show the chiral susceptibility $\chi_{\sigma}$ at $\beta = 4.0$
as a function of temperature $T$ for various chemical potential $\mu$.
The peak position of the $\chi_{\sigma}$ locates a chiral crossover and a critical endpoint (CEP).
As $\mu$ increases from zero, the peak becomes gradually larger and moves to the smaller $T$ direction.
When the $\mu$ reaches around $0.6$, the susceptibility shows a drastic enhancement,
which indicates a critical phenomena associated with the CEP.
At larger $\mu$, say $0.7$, the system is in the high-density phase where a peak is not seen for any $T$.

\begin{figure}[htb]
\includegraphics[width=8.0cm]{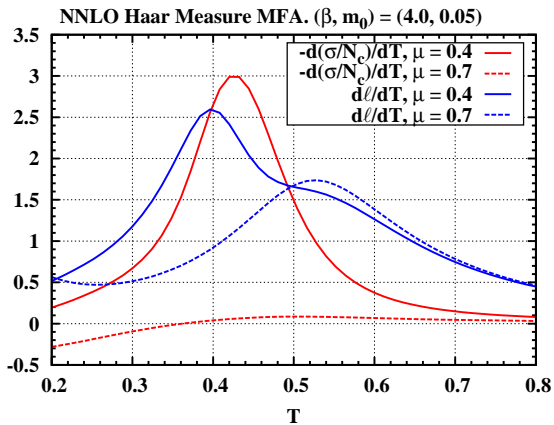}
\caption{(Color online)
The temperature derivative
of the chiral condensates and Polyakov loops in the NNLO Haar measure MFA with $m_0 = 0.05$
as a function of temperature at $\mu=0.4$ and $0.7$.
All quantities are in lattice units at the coupling $\beta = 4.0$.}
\label{Fig:peak-NNLO}
\end{figure}
In Fig.~\ref{Fig:peak-NNLO}, we show the temperature derivative 
of the chiral condensate and Polyakov loop
as a function of temperature at $\mu=0.4$ and $0.7$.
The lattice coupling $\beta$ is fixed at $4.0$.
At $\mu=0.4$ (solid lines), the chiral and deconfinement crossovers almost simultaneously take place
as indicated by their peak positions.
This property has been observed at $\mu = 0$ \cite{PNNLO-T}.
Our finding here is that the locking of the chiral and deconfinement crossovers
remains intact at finite $\mu$ as long as the spontaneous symmetry breaking exists.
At $\mu = 0.7$,
$d(\sigma/N_c)/dT$ (red-dashed line) shows no signal at any $T$ due to the absence of the chiral crossover,
and the $d\ell/dT$ (blue-dashed line) tends to lose peaklike structure.

\begin{figure*}[ht]
\includegraphics[width=8.0cm]{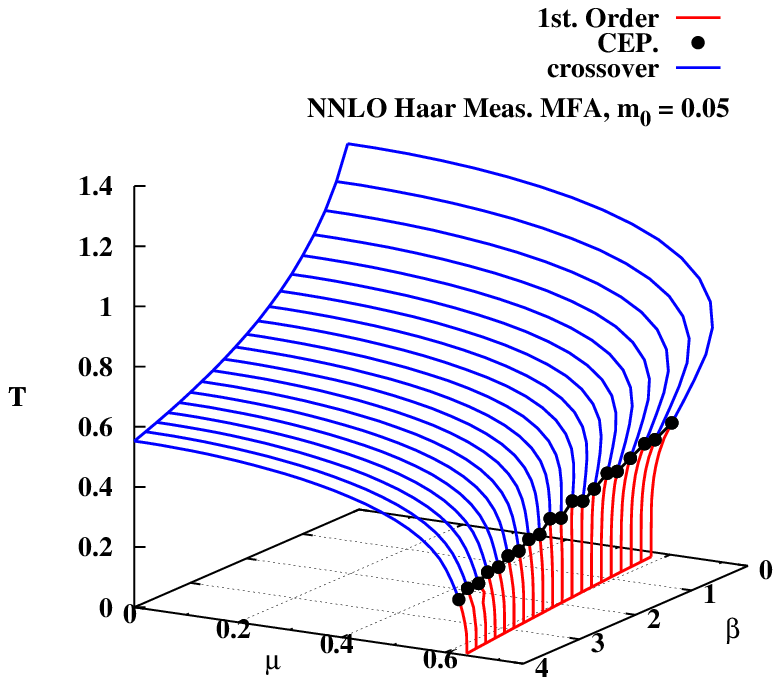}
\includegraphics[width=8.0cm]{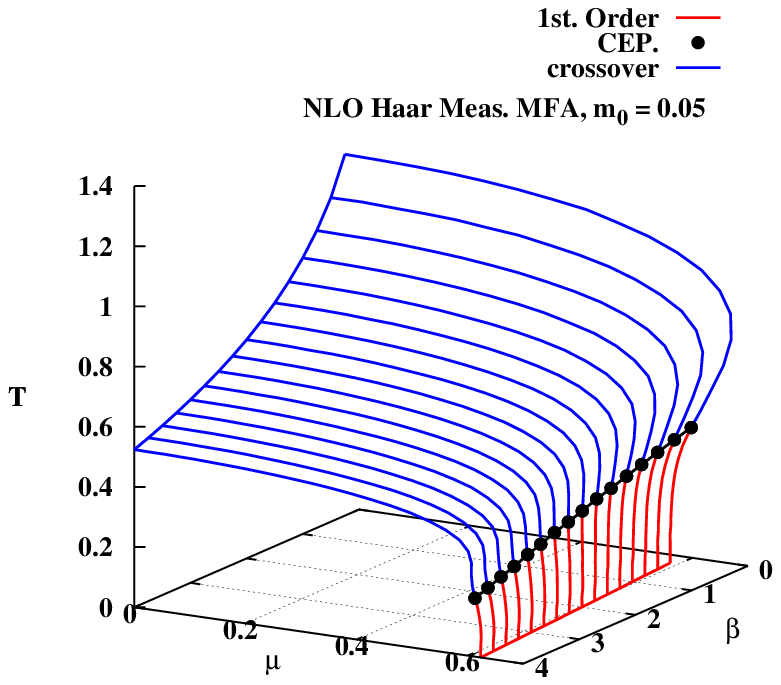}
\caption{(Color online)
The phase diagram evolution for $m_0 = 0.05$
as a function of $\beta$ in NNLO Haar-measure MFA (left),
which is compared with the counterpart in NLO Haar-measure MFA (right) with same parameters.
The red and blue lines represent the first-order chiral transition and the chiral crossover, respectively.
The black points show the CEP separating the first-order and crossover phase boundaries.
The little fluctuations of the CEP in the NNLO case are due to limited precision
in the numerical search of the maximum of chiral susceptibilities as a function of T and mu.
All quantities are in lattice units at a given $\beta$.
}\label{Fig:PDevol-NNLO}
\end{figure*}
In our previous studies ~\cite{PNNLO-T},
we have shown that the NNLO effects to the chiral phase transition/crossover at $\mu = 0$ are very small. 
We shall now investigate the impact of the NNLO effects to the phase diagram including finite $\mu$.
In the left panel of Fig.~\ref{Fig:PDevol-NNLO},
we show the phase diagram evolution as a function of $\beta$ in NNLO Haar measure MFA.
The black points represent the CEP
which separates the chiral crossover region (higher $T$, blue-solid lines)
and the first-order transition region (lower $T$, red-solid lines).
Due to the finite coupling effects,
the crossover line and the critical point move in the lower $T$ direction.
For comparison, we show the counterpart at NLO with $m_0 = 0.05$ in the right panel.
It is seen that the NNLO phase diagram (left) is very close to the NLO one (right).

In the end of the subsection \ref{subsec:NLO_Haar},
we have mentioned that the NNLO effects for the quark sector should be included,
particularly in the Haar measure MFA,
to be consistent with the plaquette-driven Polyakov loop sector
with respect to the order counting of the strong-coupling expansion.
However, the results in this subsection indicate that
the NNLO corrections are tiny in whole region of the phase diagram.
Thus, the NLO results shown in the previous subsections would be reliable.

\section{Summary}\label{sec:summary}
We have investigated the QCD phase diagram
in color SU($N_c=3$) gauge group
at finite temperature $T$ and quark chemical potential $\mu$
by using the strong-coupling expansion of the lattice QCD (SC-LQCD)
with one species of staggered fermion.
Our effective potential \cite{PNNLO-T}
includes the LO [${\cal O}(1/g^0)$], NLO [${\cal O}(1/g^2)$], 
and NNLO [${\cal O}(1/g^4)$] effects of the strong-coupling expansion
in the quark sector, and the LO effects of Polyakov loop
$\mathcal{O}([1/g^2]^{1/T})$ in the pure gluonic sector.
The Polyakov loops are
evaluated in two approximation schemes;
a simple mean-field treatment (Haar measure mean-field approximation (MFA))
and an improved treatment with fluctuation effects (Weiss MFA).
In this setup, we have investigated the whole structure of
the SC-LQCD phase diagram with a special emphasis on the Polyakov loops effects.

In both Haar measure and Weiss MFA schemes,
the first-order chiral phase boundary emerges in the low $T$ region
and ends up with the tricritical point (TCP),
from which the second-order chiral phase boundary
evolves to the smaller $\mu$ direction
with increasing $T$ in the chiral limit ($m_0=0$).
The Polyakov loop together with finite $\beta$ effects
strongly suppresses the critical temperature $T_c$
in the second-order/crossover region at small $\mu$,
while it gives a minor modification
of the first-order phase boundary at larger $\mu$.
As a result, the chiral phase boundary becomes much closer to
the expected one in the real-life QCD
as summarized in Fig.~\ref{Fig:PD_evol_00} (NLO case)
and Fig.\ref{Fig:PDevol-NNLO} (left: NNLO, right NLO).
It is also remarkable that the NNLO effects are subdominant
in whole region of the phase diagram.

In both Haar measure MFA and Weiss MFAs,
the critical point (CP) tends to go into
low $T$ region with increasing $\beta$,
and the second-order chiral phase boundary becomes dominant.
This trend is also reported in the MDP simulations
\cite{MDP-NLO,deForcrand:2014tha}
and supports the recent MC results
based on the critical surface analysis \cite{deForcrand:2008vr}.
However, the trend is opposite to
the anomaly based expectation for $N_f = 4 > 2$ \cite{Pisarski:1983ms}.
The anomaly effects in the staggered fermion formalism
should be further investigated in the future.

We have investigated thermodynamic quantities,
which is of great interest in the study of EOS of quark matter,
which has however been challenging in SC-LQCD.
Our findings are that
a pressure and an interaction measure are drastically enhanced
by Polyakov loop thermal excitations.

We have found some characteristic features of Polyakov loops at finite $\mu$.
At finite $\mu$ in the broken phase,
the anti-Polyakov loop $\bar{\ell}$ becomes larger than $\ell$,
which is interpreted as a screening effect
of quarks at equilibrium with net quark number density.
In the chirally symmetric high density phase,
the Polyakov loop becomes relatively large even at a small temperature,
which can be understood from the absence of the dynamical quark mass in the symmetric phase.

We have shown that
the chiral and Polyakov loop susceptibilities $(\chi_{\sigma},\chi_{\ell})$
have their peaks near to each other 
in the second-order transition or crossover region.
In the vicinity of the critical point,
the peak of the $\chi_{\ell}$ rapidly diminishes.
We have found two qualitative differences
between the Weiss and Haar measure MFA
on the Polyakov loop susceptibilities:
First, the peak of $\chi_{\ell}$
is more strongly locked to the chiral phase boundary in Weiss MFA
than the Haar measure MFA case.
Second, the $Z_3$ deconfinement dynamics 
artificially remains in the Haar measure MFA
and disappears by taking account of the Polyakov loop fluctuations in Weiss MFA.
Our findings are summarized in Fig.~\ref{Fig:PD_H_40_00}
(upper, Haar measure MFA result)
and \ref{Fig:PD_W_40_00} (Weiss MFA result).
The above difference results from the fact that
the effective potential of Weiss MFA does not admit any remnant of the $Z_{3}$ symmetric structure
in sharp contrast to the Haar measure MFA
and many other chiral effective models~\cite{Fukushima:2003fw,Fukushima:2008wg,Herbst:2010rf}.
Thus, the Weiss MFA does not support the isolated deconfinement transition/crossover
from the chiral phase boundary at large $\mu$.

There are several future directions to be investigated.
First, it is important to evaluate
the higher order terms of the strong-coupling expansion,
and/or to invent a resummation technique to account for the higher orders.
From this viewpoint, we find recent developments for
the Polyakov loop effective potential \cite{Langelage}.
Second, it is desirable to establish
the exact evaluation of each order of the strong-coupling expansion
beyond the mean-field approximation and $1/d$ expansion.
This will be achieved by extending
the MDP works \cite{MDP-NLO,deForcrand:2014tha}
to include the higher-order of expansions
as well as the Polyakov loop effects.
Another method to go beyond MFA is
the Monte-Carlo simulations for the auxiliary field integrals
at each order of the expansion \cite{SC-LQCD-MC}.
Third, it is interesting to evaluate
the complex phase effect of Polyakov loops;
The susceptibilities associated with the phase
may give a new probe of the QCD phase transition \cite{Lo:2013etb}.
And finally, the Weiss MFA results, 
especially the quark and Polyakov loop
thermal excitations summarized in Table \ref{Tab:FDetW},
may open a possibility to invent
an upgraded version of the PNJL-type model
which more reasonably describes the interplay between
the chiral and deconfinement dynamics.

\section*{Acknowledgments}
We thank Maria Paola Lombardo and Philippe de Forcrand for fruitful discussions.
This work was supported in part by the Grants-in-Aid
for Scientific Research from Japan Society for the Promotion of Science (JSPS)
(Nos. 22-3314,
15K05079, 
15H03663, 
16K05350), 
for Young Scientists (B) No.15K17644 (Kohtaroh Miura),
the Grants-in-Aid for Scientific Research on Innovative Areas from
Ministry of Education, Culture, Sports, Science and Technology (MEXT)
(Nos. 24105001, 24105008),
and the Yukawa International Program for Quark-hadron Sciences (YIPQS).
Kohtaroh Miura was supported by in part
by the OCEVU Laboratoire d’excellence (ANR-11-LABX-0060) and
the A*MIDEX Project (ANR-11-IDEX-0001-02), which are funded
by the ``Investissements d’Avenir'' French government
program and managed by the ``Agence nationale de la recherche'' (ANR).
\appendix
\section{Effective potential in strong-coupling lattice QCD}\label{app:Feff}
We briefly review the derivation of the effective potential Eq.~(\ref{Eq:Feff})
based on our previous papers~\cite{PNNLO-T,Nakano:2009bf}.
We start from the lattice QCD action
with one species of staggered fermion ($\chi$)
with a current quark mass ($m_0$) and chemical potential ($\mu$),
\begin{align}
\mathcal{Z}_{\mathrm{LQCD}}
&= \int\mathcal{D}[\chi,\bar{\chi},U_{\nu}]~
e^{-S_{\mathrm{LQCD}}[\chi,\bar{\chi},U_{\nu}]}
\ ,\label{Eq:S_LQCD}\\
S_{\mathrm{LQCD}}
&= S_F + S_G + m_0\sum_x\bar{\chi}_x\chi_x
\ ,\label{Eq:Z}
\end{align}
where,
\begin{align}
S_F
&= \frac{1}{2}\sum_{\nu,x}\Bigl[
\eta_{\nu,x}\bar{\chi}_x U_{\nu,x} \chi_{x+\hat{\nu}}
-\eta_{\nu,x}^{-1}(h.c.)
\Bigr]\ ,\label{Eq:SF}\\
\eta_{\nu,x}
&= \exp(\mu\,\delta_{\nu 0})(-1)^{x_0+\cdots +x_{\nu-1}}\ ,\\
S_G
&= \beta\sum_{P}
\biggl[
1-\frac{1}{2N_c}\Bigl[U_P+U_P^{\dagger}\Bigr]
\biggr]\label{Eq:SG}\ .
\end{align}
We have employed lattice units $a=1$.
The $U_{\nu,x}\in SU(N_c)$ and
$U_{P=\mu\nu,x}=\mathrm{tr}_c[U_{\mu,x}U_{\nu,x+\hat{\mu}}
U^\dagger_{\mu,x+\hat{\nu}}U^\dagger_{\nu,x}]$
represent the link- and plaquette-variable, respectively.
In the chiral limit ($m_0\to 0$),
the action has the $U_{\chi}(1)$ chiral symmetry,
which is enhanced to $SU(N_f=4)$ in the continuum limit.

\begin{table}
\caption{The hadronic composites which appears
after the spatial link integrals. 
}\label{Tab:composites}
\begin{center}
\begin{tabular}{cc}
\hline\hline
Symbol
&Composites \\
\hline
$M_x$ &
$\chibar_x \chi_x$ \\
$\bigl(\Vp_x,~\Vm_x\bigr)$ &
$\bigl(
	\bar{\chi}_x e^{\mu} U_{0,x} \chi_{x + \hat{0}},~
	\chibar_{x+\hat{0}} e^{-\mu} U_{0,x}^{\dagger} \chi_x
\bigr)$ \\
$\bigl(W^{+}_{x},W^{-}_{x}\bigr)$ &
$\bigl(
	\bar{\chi}_x
	e^{2\mu}
	U_{0,x}
	U_{0,x+\hat{0}}
	\chi_{x+2\hat{0}},~
	\bar{\chi}_{x+2\hat{0}}
	e^{-2\mu}
	U_{0,x+\hat{0}}^{\dagger}
	U_{0,x}^{\dagger}
	\chi_{x}
\bigr)$ \\
$L_{p,\mathbf{x}}$ &
$\mathrm{tr}_c\bigl[\prod_{\tau}U_{0,\mathbf{x}\tau}\bigr]/N_c$ \\
\hline\hline
\end{tabular}
\end{center}
\end{table}

\begin{table}
\caption{The auxiliary field $\Phi$ and $(\ell,\bar{\ell})$
See also Table \protect\ref{Tab:composites}.
}\label{Tab:aux}
\begin{center}
\begin{tabular}{cc}
\hline\hline
Symbol&
Mean Fields Contents \\
\hline
$\sigma$ &
$ -\langle M\rangle $ \\
$\bigl(\psibarTT,\psiTT\bigr)$ &
$\bigl(\langle W^+\rangle,~\langle W^-\rangle\bigr)$ \\
$\bigl(\psibarSS,\psiSS\bigr)$ &
$\bigl(\langle MM\rangle,~\langle MMMM\rangle\bigr)$ \\
$\bigl(\psibarTS,\psiTS\bigr)$ &
$\bigl(-\langle \Vp\Vm\rangle,~2\langle MM\rangle\bigr)$ \\
$\bigl(\psibarT,\psiT\bigr)$ &
$\bigl(-\langle \Vp\rangle,~\langle \Vm\rangle\bigr)$ \\
$\bigl(\psibarS,\psiS\bigr)$ &
$\bigl(\langle MM\rangle,~\langle MM\rangle\bigr)$ \\
\hline
$(\ell,\bar{\ell})$ &
$(\langle L_{p}\rangle,\langle\bar{L}_{p}\rangle)$ \\
\hline\hline
\end{tabular}
\end{center}
\end{table}
\begin{table}
\caption{The coupling coefficients
appearing in the effective action/potential.
Here, $g$, $N_c=3$, and $d=3$ represents
the gauge coupling, number of color, and spatial dimension,
respectively.
See Table \protect\ref{Tab:aux} for the auxiliary fields
$(\psi_{\cdots},\bar{\psi}_{\cdots})$.
}\label{Tab:coupling}
\begin{center}
\begin{tabular}{cc}
\hline\hline
Symbol &
Definition \\
\hline
$b_{\sigma}$&
$d/(2N_c)$\\
$\beta_t$&
$\bigl(d/(N_c^2g^2)\bigr)\cdot \bigl(1+1/(2g^2)\bigr)$\\
$\beta_s$&
$\bigl(d(d-1)/(8N_c^4g^2)\bigr)\cdot \bigl(1+1/(2g^2)\bigr)$\\
$\bsigp$&
$b_{\sigma} + 2 \bigl[ \beta_{ss}\psi_{ss}
+\beta_{\tau s}\bar{\psi}_{\tau s}
+\beta_s^{\prime}(\psiS+\psibarS)\bigr]$ \\
$\beta_{t}^{\prime}$&
$\beta_t
+\beta_{\tau s}\psi_{\tau s}$\\
$\beta_{s}^{\prime}$&
$\beta_s
+2\beta_{ss}\bar{\psi}_{ss}$\\
$\beta_{\tau\tau}$&
$d/(2N_c^3g^4)$\\
$\beta_{ss}$&
$d(d-1)(d-2)/(16N_c^7g^4)$\\
$\beta_{\tau s}$&
$d(d-1)/(2N_c^5g^4)$\\
\hline\hline
\end{tabular}
\end{center}
\end{table}

\begin{table}
\caption{
Quantities which govern the property of the effective potential.
See Table \protect\ref{Tab:coupling} for the couplings
$(\bsigp,\bt^{\prime},\btt)$
and Table \protect\ref{Tab:aux} for the auxiliary fields
$(\sigma,\psiT,\psibarT,\psiTT,\psibarTT)$.
}\label{Tab:renorm}
\begin{center}
\begin{tabular}{c|c|c}
\hline\hline
Symbol & Definition & Meanings \\
\hline
$\tilde{m}_q$ & $m_q^{\prime}/\sqrt{Z_+Z_-}$ & dynamical quark mass \\
& $m_q^{\prime}=\bsigp \sigma + m_0 $ & \\
& $\quad - \btt(\psibarTT+\psiTT)$ & \\
\hline
$\sqrt{Z_+Z_-}$ & $\Zp=1+\bt'\psibarT$ & wave function\\
& $\quad +4\btt m_q^{\prime}\psibarTT$ & renormalization factor\\
& $\Zm=1+\bt'\psiT$ & \\
& $\quad +4\btt m_q^{\prime}\psiTT$ & \\
\hline
$E_q$ & $\sinh^{-1} \tilde{m}_q$ & quark excitation energy\\
\hline
$\tilde{\mu}$ & $\mu - \log\sqrt{Z_+/Z_-}$ & shifted chemical potential\\
\hline\hline
\end{tabular}
\end{center}
\end{table}

There are four main steps to derive the effective potential
from the lattice QCD action~(\ref{Eq:S_LQCD})~\cite{PNNLO-T}:
First, we carry out the strong-coupling expansion,
and integrate out the spatial link variables in each order.
The effective action is obtained as a function of various hadronic composites.
For the composites including
the staggered quarks ($\chi,\bar{\chi}$),
we take account of the terms up to ${\cal O}(1/g^6)$,
and extract from them the leading order terms of the $1/d$ expansion
$\mathcal{O}(1/d^0)$~\cite{KlubergStern:1982bs}.
For the pure gluonic composites,
we take account of
the leading order contributions to the Polyakov-loop
[$\mathcal{O}(1/g^{2N_\tau}),~N_\tau:$ lattice temporal extension].
The hadronic composites are summarized in Table \ref{Tab:composites},
and the effective action is expressed by using these composites,
\begin{align}
S_{\mathrm{eff}} = 
S_{\mathrm{eff}}^{\sub{NNLO}} + S_{\mathrm{eff}}^{\mathrm{Pol}}
\ ,\label{Eq:Seff_app}
\end{align}
with
\begin{align}
&S_{\mathrm{eff}}^{\mathrm{Pol}} =
-N_c^2\biggl(\frac{1}{g^2N_c}\biggr)^{N_{\tau}=1/T}
\sum_{j,\mathbf{x}}\Bigl[
\bar{L}_{p,\mathbf{x}}L_{p,\mathbf{x}+\hat{j}}+h.c.
\Bigr]\ ,\label{Eq:Seff_Pol_app}
\end{align}
and
\begin{align}
&S_{\mathrm{eff}}^{\sub{NNLO}} =
\sum_x\frac12 (V^+_x - V^-_x)
+\sum_{x,j>0}\Biggl[- \frac{b_\sigma}{2d} [MM]_{j,x}
\nn\\
&+\frac{\bt}{4d} [V^+V^- + V^-V^+]_{j,x}
- \sum_{k>0,k\not=j} \frac{\beta_s [MMMM]_{jk,x}}{2d(d-1)}\nn\\
&-\frac {\beta_{\tau\tau}}{2d}[W^+W^- + W^-W^+]_{j,x}
\nn\\
& +\sum_{|k| \neq j}
\biggl[
\sum_{\substack{|k|,|l|>0,\\|l|\neq j,|l|\neq|k|}}
\frac {-\beta_{ss}[MMMM]_{jk,x}[MM]_{j,x+\hat{l}}}{4d(d-1)(d-2)}\nn\\
&+ \frac{
\beta_{\tau s}
[V^+V^- + V^-V^+]_{j,x}
}{8d(d-1)}\nn\\
&\qquad\times
\bigl( [MM]_{j,x+\hat{k}} + [MM]_{j,x+\hat{k}+\hat{0}} \bigr)
\biggr]
\Biggl]\ .\label{Eq:Seff_NNLO_app}
\end{align}
We have introduced a short-hand notation
\begin{align}
&[AB]_{j,x} = A_xB_{x+\hat{j}}\ ,\\
&[ABCD]_{jk,x}= A_xB_{x+\hat{j}}C_{x+\hat{j}+\hat{k}}D_{x + \hat{k}}\ ,
\end{align}
and the couplings $\beta_{\cdots}$ in Eq.~(\ref{Eq:Seff_NNLO_app})
are summarized in Table \ref{Tab:coupling}.

\begin{table*}[htb]
\caption{The thermal excitation effects
$\mathcal{P}^I_n$ and $\mathcal{Q}^I$
in the quark determinant of the Weiss MFA,
Eq.~(\protect\ref{Eq:FDetW_app}).
The left column represents the excitation channel
with the label $I$ in the text:
$(\mathrm{M},\mathrm{B},\mathrm{Q},\mathrm{D})$
stands for (mesonic,baryonic,quark,diquark) excitation.
The quark excitation energy $E_q$
and modified chemical potential $\tilde{\mu}$
appearing in the third column
are explained in Table\protect\ref{Tab:renorm}.
In the right column,
$I_n$ represents a modified Bessel function
with the argument
$x=4dN_c\bigl(\beta/(2N_c^2)\bigr)^{1/T}\sqrt{\ell\bar{\ell}}$.
}\label{Tab:FDetW}
\begin{center}
\begin{tabular}{l|l|l|l}
\hline\hline
Excitation ($I$) &
$N_{\mathrm{Q}}^{I}$ &
$\mathcal{Q}^{I}(\Phi)$ &
$\mathcal{P}^I_n\Bigl(\sqrt{\ell\bar{\ell}}\Bigr)$ \\
\hline
$\mathrm{MMM}$&
$0$&
$\Bigl(2\cosh(E_q/T)\Bigr)^{N_c}$&
$\mathcal{P}_n^{~\sub{MMM}} = I_n^3-I_{n-2}I_nI_{n+2}-2I_{n-1}I_nI_{n+1}+I_{n-2}I_{n+1}^2+I_{n-1}^2I_{n+2}$\\
\hline
$\mathrm{MQ\bar{Q}}$&
$0$&
$2\cosh(E_q/T)$&
$\mathcal{P}_n^{~\sub{MQ\bar{Q}}} = -2(I_n^3-I_{n-2}I_nI_{n+2})+5I_{n-1}I_nI_{n+1}-3(I_{n-2}I_{n+1}^2+I_{n-1}^2I_{n+2})$\\
\quad&
\quad&
\quad&
\qquad $-I_{n-3}I_nI_{n+3}+I_{n-1}I_{n-2}I_{n+3}+I_{n-3}I_{n+1}I_{n+2}$\\
\hline
$\mathrm{B}$&
$3$&
$e^{N_c\tilde{\mu}/T}$&
$\mathcal{P}_n^{~\sub{B}}=\mathcal{P}_{n-1}^{~\sub{MMM}}$\\
\hline
$\mathrm{\bar{B}}$&
$-3$&
$e^{-N_c\tilde{\mu}/T}$&
$\mathcal{P}_n^{~\sub{\bar{B}}}=\mathcal{P}_{n+1}^{~\sub{MMM}}$\\
\hline
$\mathrm{MMQ}$&
$1$&
$e^{\tilde{\mu}/T}\Bigl(2\cosh(E_q/T)\Bigr)^{2}$&
$\mathcal{P}_n^{~\sub{MMQ}} = I_{n-1}I_n^2+I_{n-3}I_{n+1}^2-I_{n-1}^2I_{n+1}+I_{n-2}I_{n-1}I_{n+2}-I_{n-3}I_{n}I_{n+2}-I_{n-2}I_nI_{n+1}$\\
\hline
$\mathrm{MM\bar{Q}}$&
$-1$&
$e^{-\tilde{\mu}/T}\Bigl(2\cosh(E_q/T)\Bigr)^{2}$&
$\mathcal{P}_n^{~\sub{MM\bar{Q}}} = I_{n+1}I_n^2+I_{n+3}I_{n-1}^2-I_{n+1}^2I_{n-1}+I_{n-2}I_{n+1}I_{n+2}-I_{n+3}I_{n}I_{n-2}-I_{n+2}I_nI_{n-1}$\\
\hline
$\mathrm{MD}$&
$2$&
$e^{2\tilde{\mu}/T}~2\cosh(E_q/T)$&
$\mathcal{P}_n^{~\sub{MD}}=\mathcal{P}_{n-1}^{~\sub{MM\bar{Q}}}$\\
\hline
$\mathrm{M\bar{D}}$&
$-2$&
$e^{-2\tilde{\mu}/T}~2\cosh(E_q/T)$&
$\mathcal{P}_n^{~\sub{M\bar{D}}}=\mathcal{P}_{n+1}^{~\sub{MMQ}}$\\
\hline
$\mathrm{D\bar{Q}}$&
$1$&
$e^{\tilde{\mu}/T}$&
$\mathcal{P}_n^{~\sub{D\bar{Q}}} = 2(I_{n-1}^2I_{n+1}-I_{n-2}I_{n-1}I_{n+2}+I_{n-3}I_{n}I_{n+2})$\\
\quad&
\quad&
\quad&
\qquad $+I_{n-2}^2I_{n+3}-I_{n-1}I_n^2-I_{n-3}I_{n+1}^2-I_{n-3}I_{n-1}I_{n+3}$\\
\hline
$\mathrm{Q\bar{D}}$&
$-1$&
$e^{-\tilde{\mu}/T}$&
$\mathcal{P}_n^{~\sub{Q\bar{D}}} = 2(I_{n+1}^2I_{n-1}-I_{n-2}I_{n+1}I_{n+2}+I_{n-2}I_{n}I_{n+3})$\\
\quad&
\quad&
\quad&
\qquad $+I_{n+2}^2I_{n-3}-I_{n+1}I_n^2-I_{n+3}I_{n-1}^2-I_{n-3}I_{n+1}I_{n+3}$\\
\hline\hline
\end{tabular}
\end{center}
\end{table*}

Second, we introduce the auxiliary fields
for the hadronic composites
to bosonize the effective action $S_{\mathrm{eff}}^{\sub{NNLO}}$,
and perform the static mean-field and saddle-point approximations.
The auxiliary fields are summarized in Table \ref{Tab:aux},
and the $S_{\mathrm{eff}}^{\sub{NNLO}}$ reduces into
\begin{align}
S_{\mathrm{eff}}^{\sub{NNLO}}\simeq
S_{\mathrm{eff}}^{\sub{F}} + S_{\mathrm{eff}}^{\sub{X}}
\ ,\label{Eq:Seff_NNLO_app_red}
\end{align}
where
\begin{align}
&S_{\mathrm{eff}}^{\sub{F}} =
\sqrt{Z_+Z_-}
\sum_{xy}\bar{\chi}_x G_{xy}^{-1}(\tilde{m}_q,\tilde{\mu})\chi_y
\ ,\label{Eq:Seff_F_app}\\
& G_{xy}^{-1}(\tilde{m}_q,\tilde{\mu}) \nn\\
&= \tilde{m}_{q}\delta_{xy}
+\frac{\delta_{\mathbf{xy}}}{2}
\Bigl[
e^{\tilde{\mu}}U_{0,x}\delta_{x+\hat{0},y}
-e^{-\tilde{\mu}}U_{0,x}^{\dagger}\delta_{x-\hat{0},y}
\Bigr]\ , \label{Eq:qhop}\ \\
&S_{\mathrm{eff}}^{\sub{X}}=
N_{\tau}N_s^d
\biggl[
\bsigp \sigma^2
+\frac{1}{2}\bt'\bar{\psi}_\tau\psi_\tau
+\frac{1}{2}\beta_s'\psibarS\psiS \nn\\
&+\beta_{\tau\tau}\psibarTT\psiTT
+\beta_{ss}\bar{\psi}_{ss}\psi_{ss}
+\frac{1}{2}\beta_{\tau s}\bar{\psi}_{\tau s}\psi_{\tau s}
\biggr]\ .\label{Eq:Seff_X_app}
\end{align}
Here, the dynamical quark mass $\tilde{m}_q$,
the shifted quark chemical potential $\tilde{\mu}$,
and the wave function renormalization factor $\sqrt{Z_+Z_-}$
are summarized in Table \ref{Tab:renorm},
and the $N_{t(s)}$ represents the temporal (spatial) lattice extension.

Third, we carry out the Gaussian integral over the staggered quarks
$(\chi,\bar{\chi})$ in Eq.~(\ref{Eq:Seff_F_app})
in the antiperiodic boundary condition.
The resultant quark determinant at finite $T$ is then
calculated by using the Matsubara method
in the Polyakov gauge for temporal link variables~\cite{DKS},
\begin{align}
&\int\mathrm{D}[\chi,\bar{\chi}]~e^{-S_{\mathrm{eff}}^{\sub{F}}}
= \prod_{\mathbf{x}}
\biggl[
 e^{N_c(\log\sqrt{Z_+Z_-}+E_q)/T}
\nn\\
&\times \mathrm{det}_c \Bigl[ 
 \bigl(1+N_cL_{p,\mathbf{x}} e^{-(E_q-\tilde{\mu})/T}\bigr)\nn\\
&\qquad
 \bigl(1+N_c\bar{L}_{p,\mathbf{x}} e^{-(E_q+\tilde{\mu})/T}\bigr)
 \Bigr]
\biggl]\ ,\label{Eq:eVq}
\end{align}
with $E_q = \sinh^{-1}\tilde{m}_q$.
Temperature $T$ is now considered as a continuous valued number
(see the appendix in Ref.~\cite{Kawamoto:2005mq} for details).
The Polyakov loop $L_{p,\mathbf{x}}$ has appeared in the determinant
via the quark hopping wrapping around the temporal direction
in addition to the Plaquette effects Eq.~(\ref{Eq:Seff_Pol_app}).

Finally, we evaluate the $L_{p,\mathbf{x}}$ effects
in the path integral over the temporal link variable $U_0$
in two approximation schemes:
Haar measure and Weiss MFA.
In the former, we replace the Polyakov loop $L_{p,\mathbf{x}}$
contained in Eq.~(\ref{Eq:Seff_Pol_app}) and (\ref{Eq:eVq}) as well as
the Haar measure of the $U_0$ path integral
with a constant mean-field $(\ell,\bar{\ell})$
instead of performing the $U_0$ path integral.
In the latter, 
we introduce a mean-field $(\ell,\bar{\ell})$
via the extended Hubbard-Stratonovich
transformation~\cite{NLO} in Eq.~(\ref{Eq:Seff_Pol_app}), 
and exactly carry out the $U_0$ path integral
to include the fluctuation effects from $(\ell,\bar{\ell})$~\cite{PNNLO-T}.

As a result, we obtain the effective potential
\begin{align}
&\mathcal{F}_{\mathrm{eff}}^{\sub{H/W}}(\Phi,\ell,\bar{\ell};\beta,m_0,T,\mu)\nn\\
&=\mathcal{F}_{\sub{X}}(\Phi,\beta)
+\mathcal{F}_{\mathrm{det}}^{\sub{H/W}}(\Phi,\beta,m_0,T,\mu)\nn\\
&\qquad
+\mathcal{F}_{\sub{P}}^{\sub{H/W}}(\ell,\bar{\ell},\beta,T)
+\mathcal{O}(1/g^6,1/\sqrt{d})
\ .\label{Eq:Feff_app}
\end{align}
The auxiliary field term is 
given by Eq.~(\ref{Eq:Seff_X_app})
and common in both Haar measure MFA and Weiss MFA,
\begin{align}
&\mathcal{F}_{\sub{X}}(\Phi,\beta)
=S_{\mathrm{eff}}^{\sub{X}}/(N_tN_s^3)
\ .\label{Eq:Faux_app}
\end{align}
The quark determinant and the Polyakov loop effects are given as
\begin{align}
&\mathcal{F}_{\mathrm{det}}^{\sub{H}}
=-N_cE_q- N_c \log \sqrt{Z_+Z_-}\nn\\
&\quad
-T\Bigl(\log \mathcal{R}_{q}(E_q-\tilmu,\ell,\lbar)
+\log \mathcal{R}_q(E_q+\tilmu,\lbar, \ell)
\Bigr)\ ,\label{Eq:FDet_Haar_app}\\
&\mathcal{R}_{q}(x,y,\bar{y})
\equiv 1+N_c(y e^{-x/T} + \bar{y} e^{-2x/T})+e^{-3x/T}\nn\\
&\mathcal{F}_{\sub{P}}^{\sub{H}}
=-2TdN_c^2\biggl(\frac{1}{g^2N_c}\biggr)^{1/T}\bar{\ell}\ell
-T\log \mathcal{R}_{\mathrm{Haar}}(\ell,\bar{\ell})\ ,\label{Eq:FPol_Haar_app}\\
&\mathcal{R}_{\mathrm{Haar}}(\ell,\bar{\ell})
\equiv
1-6\bar{\ell}\ell
-3\bigl(\bar{\ell}\ell\bigr)^2
+4\bigl(\ell^{N_c}+\bar{\ell}^{N_c}\bigr)
\ ,\label{Eq:R_Haar_app}
\end{align}
in Haar measure MFA case, and
\begin{align}
&\mathcal{F}_{\sub{P}}^{\sub{W}}
+\mathcal{F}_{\mathrm{det}}^{\sub{W}}
= 2TdN_c^2\biggl(\frac{1}{g^2N_c}\biggr)^{1/T}
\bar{\ell}\ell\nn\\
&\quad -
T\log\biggl[
\sum_{I}
\mathcal{Q}^{I}(\Phi)
\mathcal{P}^{I}(\ell,\bar{\ell})~
\biggr]\ ,\label{Eq:FDetW_app}\\
&\mathcal{P}^{I}(\ell,\bar{\ell})
=
\sum_{n=-\infty}^{\infty}
\biggl(
\sqrt{\ell/\bar{\ell}}
\biggr)^{-N_cn + N_{\mathrm{Q}}^I}
\mathcal{P}^{I}_n\biggl(\sqrt{\ell\bar{\ell}}\biggr)
\ ,\label{Eq:Pn_app}
\end{align}
in Weiss MFA case.
In Eqs.~(\ref{Eq:FDetW_app}) and (\ref{Eq:Pn_app}),
the index $I$ labels
a pattern of thermal excitations of the quark composites,
and the fermionic thermal excitation effects
$\mathcal{Q}^{I}$,
the Polyakov loop thermal excitation effects
$\mathcal{P}^{I}_n$,
and the quark number index $N_{\mathrm{Q}}^{I}$
are summarized in Table \ref{Tab:FDetW}.

As indicated in Eq.~(\ref{Eq:FPol_Haar_app}) and (\ref{Eq:FDetW_app}),
the $Z_3$ symmetric term remains in the Haar measure MFA,
but not in the Weiss MFA up to the first $\bar{\ell}\ell$ term.
In the latter,
the path integral over the temporal link variable $U_0$
which accounts for the summation over the Polyakov loop fluctuations
spoils the $Z_3$ symmetry in the presence of the dynamical quarks.
In heavy quark mass limit $m_0 \to \infty$,
the $Z_3$ symmetry recovers in the Weiss MFA as follows:
In the effective potential of Weiss MFA,
the factor 
$
\bigl(
\sqrt{\ell/\bar{\ell}}
\bigr)^{-N_cn + N_{\mathrm{Q}}^I}
$
in Eq.~(\ref{Eq:Pn_app})
gives a unique source of
the explicit $Z_3$ symmetry breaking
($(\ell,\bar{\ell}) \to (\Omega\ell,\Omega^{-1}\bar{\ell})$, $\Omega\in Z_3$).
For $m_0 \to \infty$ or equivalently $E_q\gg T,\mu$,
the three mesonic thermal excitation
$\mathcal{Q}^{I=\mathrm{MMM}}$ in Table \ref{Tab:FDetW} becomes dominant,
and it does not carry
the quark number $N_{\mathrm{Q}}^{I=\mathrm{MMM}}=0$.
Therefore, the Eq.~(\ref{Eq:FDetW}) reduces to
\begin{align}
&\mathcal{F}_{\sub{P}}^{\sub{W}}
+\mathcal{F}_{\mathrm{det}}^{\sub{W}}
= 2TdN_c^2\biggl(\frac{1}{g^2N_c}\biggr)^{1/T}
\bar{\ell}\ell
-T\log\biggl[
\mathcal{Q}^{I=\mathrm{MMM}}(\Phi)\nn\\
&\qquad\times
\sum_{n=-\infty}^{\infty}
\biggl(
\sqrt{\ell/\bar{\ell}}
\biggr)^{-N_cn}
\mathcal{P}^{I=\mathrm{MMM}}_n\biggl(\sqrt{\ell\bar{\ell}}\biggr)
\biggr]\ .\label{Eq:FDetW_app_red}
\end{align}
This expression is invariant under the $Z_3$ transformation,
$(\ell,\bar{\ell}) \to (\Omega\ell,\Omega^{-1}\bar{\ell})$
with the property $\Omega^{N_cn}=\mathbf{1}$ for $N_c=3$.

Finally, we consider the confinement limit ($\ell,\bar{\ell}\to 0$)
in the Weiss MFA.
The quark determinant effect (\ref{Eq:FDetW_app})
includes the Polyakov loop thermal excitation $\mathcal{P}^{I}_n$,
which are solely characterized by the nth-order modified Bessel functions
as shown in Table \ref{Tab:FDetW}.
In the limit ($\ell,\bar{\ell}\to 0$),
the 0th-order modified Bessel function remains finite ($I_0(x\to 0)=1$)
while the others vanish ($I_{n\neq 0}(x\to 0)=0$).
Consequently,
the only thermal excitations which carry the quark number $0$ and $\pm 3$
survives in Table \ref{Tab:FDetW}, and the effective potential reduces into
the one which we have derived in our previous work \cite{NLO},
\begin{align}
&\mathcal{F}_{\mathrm{eff}}^{\sub{W}}
(\Phi,\ell,\bar{\ell};\beta,m_0,T,\mu)|_{\ell,\bar{\ell}=0}\to\nn\\
&\mathcal{F}_{\mathrm{eff}}^{\sub{NLO}}(\Phi;\beta,m_0,T,\mu)
=\mathcal{F}_{\sub{X}}(\Phi,\beta)\nn\\
&-T\log\biggl[
\Bigl(2\cosh\frac{E_q}{T}\Bigr)^{N_c}
-4\cosh\frac{E_q}{T}
+2\cosh\frac{N_c\tilde{\mu}}{T}
\biggr]\ .\label{Eq:app_NLO}
\end{align}





\begin{thebibliography}{99}

\bibitem{Philipsen:2012nu}
For a recent lattice review of the QCD at finite temperature and/or density, see
  O.~Philipsen,
  Prog.\ Part.\ Nucl.\ Phys.\  {\bf 70}, 55 (2013).

\bibitem{Fukushima:2013rx}
For a recent review of the QCD phase diagram, see
  K.~Fukushima and C.~Sasaki,
  Prog.\ Part.\ Nucl.\ Phys.\  {\bf 72}, 99 (2013).


\bibitem{Borsanyi:2010bp}
For recent results and reviews, see,
  S.~Borsanyi, Z.~Fodor, C.~Hoelbling, S.~D.~Katz, S.~Krieg, C.~Ratti and K.~K.~Szabo [Wuppertal-Budapest Collaboration],
  J. High Energy Phys.  {\bf 1009}, 073 (2010).


\bibitem{CSC}
D.~Bailin and A.~Love,
  Phys.\ Rept.\  {\bf 107}, 325 (1984);
R.~Rapp, T.~Schafer, E.~V.~Shuryak and M.~Velkovsky,
  Phys.\ Rev.\ Lett.\  {\bf 81}, 53 (1998); 
M.~G.~Alford, K.~Rajagopal and F.~Wilczek,
  Phys.\ Lett.\ B {\bf 422}, 247 (1998); 
M.~G.~Alford,
  Ann.\ Rev.\ Nucl.\ Part.\ Sci.\  {\bf 51}, 131 (2001);
D.~T.~Son,
  Phys.\ Rev.\ D {\bf 59}, 094019 (1999);
M.~G.~Alford, K.~Rajagopal and F.~Wilczek,
  Nucl.\ Phys.\ B {\bf 537}, 443 (1999);
J.~Berges and K.~Rajagopal,
  Nucl.\ Phys.\ B {\bf 538}, 215 (1999);
K.~Iida and G.~Baym,
  Phys.\ Rev.\ D {\bf 63}, 074018 (2001)
  [Erratum-ibid.\ D {\bf 66}, 059903 (2002)];
M.~Iwasaki and T.~Iwado,
  Phys.\ Lett.\ B {\bf 350}, 163 (1995).



\bibitem{McLerran:2007qj}
  L.~McLerran and R.~D.~Pisarski,
  Nucl.\ Phys.\  A {\bf 796}, 83 (2007).


\bibitem{QY-dev}
  Y.~Hidaka, L.~D.~McLerran and R.~D.~Pisarski,
  Nucl.\ Phys.\  A {\bf 808}, 117 (2008);
%
  L.~McLerran, K.~Redlich and C.~Sasaki,
  Nucl.\ Phys.\  A {\bf 824}, 86 (2009).


\bibitem{McLerran:2009ve}
  L.~McLerran,
  Nucl.\ Phys.\ A {\bf 830}, 709C (2009).


\bibitem{chiral-spiral}
  T.~Kojo, Y.~Hidaka, K.~Fukushima, L.~D.~McLerran and R.~D.~Pisarski,
  Nucl.\ Phys.\ A {\bf 875}, 94 (2012);
%
  T.~Kojo, Y.~Hidaka, L.~McLerran and R.~D.~Pisarski,
  Nucl.\ Phys.\ A {\bf 843}, 37 (2010).


\bibitem{Gale:2013da}
For a recent review,
see,
  C.~Gale, S.~Jeon and B.~Schenke,
  Int.\ J.\ Mod.\ Phys.\ A {\bf 28}, 1340011 (2013).


\bibitem{CEP}
  M.~Asakawa and K.~Yazaki,
  Nucl.\ Phys.\  A {\bf 504}, 668 (1989);
%
  M.~A.~Stephanov, K.~Rajagopal and E.~V.~Shuryak,
  Phys.\ Rev.\ Lett.\  {\bf 81}, 4816 (1998).


\bibitem{Mohanty:2011nm} 
  B.~Mohanty [STAR Collaboration],
  J.\ Phys.\ G {\bf 38}, 124023 (2011)
  [arXiv:1106.5902 [nucl-ex]].


\bibitem{Muroya:2003qs}
For a review of a finite chemical potential on lattice, see
  S.~Muroya, A.~Nakamura, C.~Nonaka and T.~Takaishi,
  Prog.\ Theor.\ Phys.\  {\bf 110}, 615 (2003).


\bibitem{Aarts:2013bla} 
  G.~Aarts,
  PoS LATTICE {\bf 2012}, 017 (2012)
  [arXiv:1302.3028 [hep-lat]].


\bibitem{Fujii:2013sra} 
  H.~Fujii, D.~Honda, M.~Kato, Y.~Kikukawa, S.~Komatsu and T.~Sano,
  JHEP {\bf 1310}, 147 (2013)
  [arXiv:1309.4371 [hep-lat]].

\bibitem{Cristoforetti:2012su} 
  M.~Cristoforetti {\it et al.}  [AuroraScience Collaboration],
  Phys.\ Rev.\ D {\bf 86}, 074506 (2012)
  [arXiv:1205.3996 [hep-lat]].



\bibitem{TextBook}
The review of the pioneering works for 
the strong-coupling expansion is found in the text book,
  I.~Montvay and G.~M\"unster,  
  ``Quantum Fields on a Lattice,''
  Cambridge~University~Press,~1994;
%
M.~Creutz,
``Quarks, Gluons, and Lattices,''
Cambridge Univ. Press, 1983.


\bibitem{Wilson:1974sk}
  K.~G.~Wilson,
  Phys.\ Rev.\  D {\bf 10}, 2445 (1974).


\bibitem{Creutz-MC}
  M.~Creutz,
  Phys.\ Rev.\  D {\bf 21}, 2308 (1980);
  M.~Creutz and K.~J.~M.~Moriarty,
  Phys.\ Rev.\  D {\bf 26}, 2166 (1982).


\bibitem{Munster:1980iv}
  G.~M\"unster,
  Nucl.\ Phys.\  B {\bf 180}, 23 (1981).



\bibitem{Fukushima:2003vi}
  K.~Fukushima,
  Prog.\ Theor.\ Phys.\ Suppl.\  {\bf 153}, 204 (2004).



\bibitem{Nishida:2003uj}
  Y.~Nishida, K.~Fukushima and T.~Hatsuda,
  Phys.\ Rept.\  {\bf 398}, 281 (2004).


\bibitem{Nishida:2003fb}
  Y.~Nishida,
  Phys.\ Rev.\  D {\bf 69}, 094501 (2004).


\bibitem{Azcoiti:2003eb}
  V.~Azcoiti, G.~Di Carlo, A.~Galante and V.~Laliena,
  J. High Energy Phys. {\bf 09}, 014 (2003).


\bibitem{Kawamoto:2005mq}
  N.~Kawamoto, K.~Miura, A.~Ohnishi and T.~Ohnuma,
  Phys.\ Rev.\  D {\bf 75}, 014502 (2007).


\bibitem{NLO}
  K.~Miura, T.~Z.~Nakano, A.~Ohnishi and N.~Kawamoto,
  Phys.\ Rev.\  D {\bf 80}, 074034 (2009);
%
  K.~Miura, T. Z~Nakano and A.~Ohnishi,
  Prog.\ Theor.\ Phys.\  {\bf 122}, 1045 (2009).



\bibitem{Miura:2011kq}
  K.~Miura, T.~Z.~Nakano, A.~Ohnishi and N.~Kawamoto,
  arXiv:1106.1219 [hep-lat].


\bibitem{Nakano:2009bf}
  T.~Z.~Nakano, K.~Miura and A.~Ohnishi,
  Prog.\ Theor.\ Phys.\  {\bf 123}, 825 (2010).
\bibitem{PNNLO-T}
  T.~Z.~Nakano, K.~Miura and A.~Ohnishi,
  Phys.\ Rev.\  D {\bf 83}, 016014 (2011).
%


\bibitem{Bringoltz:2006pz}
  B.~Bringoltz,
  J. High Energy Phys. {\bf 03}, 016 (2007).


\bibitem{deForcrand:2009dh}
  P.~de Forcrand and M.~Fromm,
  Phys.\ Rev.\ Lett.\  {\bf 104}, 112005 (2010).


\bibitem{deForcrand:2014tha}
  P.~de Forcrand, J.~Langelage, O.~Philipsen and W.~Unger,
  arXiv:1406.4397 [hep-lat].


\bibitem{MDP-NLO}
  P.~de Forcrand, M.~Fromm, J.~Langelage, K.~Miura, O.~Philipsen and W.~Unger,
  arXiv:1111.4677 [hep-lat];
%
  P.~de Forcrand, J.~Langelage, O.~Philipsen and W.~Unger,
  arXiv:1312.0589 [hep-lat].



\bibitem{SC-LQCD-MC}
  T.~Ichihara, A.~Ohnishi and T.~Z.~Nakano,
  PTEP {\bf 2014} (2014), 123D02
  [arXiv:1401.4647 [hep-lat]];
  T.~Ichihara, T.~Z.~Nakano and A.~Ohnishi,
  Proc.\ Sci.\ LAT2013, 143 (2014);
  A.~Ohnishi, T.~Ichihara and T.~Z.~Nakano,
  Proc.\ Sci.\ LAT2012, 088 (2012).



\bibitem{Gocksch:1984yk}
  A.~Gocksch and M.~Ogilvie,
  Phys.\ Rev.\  D {\bf 31}, 877 (1985).


\bibitem{Ilgenfritz:1984ff}
  E.~M.~Ilgenfritz and J.~Kripfganz,
  Z.\ Phys.\  C {\bf 29}, 79 (1985).

\bibitem{Fukushima:2003fm}
  K.~Fukushima,
  Phys.\ Rev.\  D {\bf 68}, 045004 (2003).


\bibitem{Fukushima:2003fw}
  K.~Fukushima,
  Phys.\ Lett.\  B {\bf 591}, 277 (2004).


\bibitem{Forcrand}
P.~de Forcrand, private communication:
For the temporal lattice extension $N_t = 2$,
the lattice bare coupling associated with the chiral phase transition
$(\beta_c)$ is found to be 3.67 at the bare quark mass $m_0 = 0.025$
and 3.81 at $m_0 = 0.05$ for one species of staggered fermion.


\bibitem{MC4}
  G.~Boyd, J.~Fingberg, F.~Karsch, L.~Karkkainen and B.~Petersson,
  Nucl.\ Phys.\  B {\bf 376}, 199 (1992);
%
  R.~V.~Gavai {\it et al.}  [MT(c) Collaboration],
  Phys.\ Lett.\  B {\bf 241}, 567 (1990);
%
 S.~A.~Gottlieb, W.~Liu, D.~Toussaint, R.~L.~Renken and R.~L.~Sugar,
 Phys.\ Rev.\  D {\bf 35}, 3972 (1987);
%
  A.~D.~Kennedy, J.~Kuti, S.~Meyer and B.~J.~Pendleton,
  Phys.\ Rev.\ Lett.\  {\bf 54}, 87 (1985).


\bibitem{Engels:1996ag}
  J.~Engels, R.~Joswig, F.~Karsch, E.~Laermann, M.~Lutgemeier and B.~Petersson,
  Phys.\ Lett.\ B {\bf 396}, 210 (1997).
%


\bibitem{V-Ploop}
  J.~Polonyi and K.~Szlachanyi,
  Phys.\ Lett.\  B {\bf 110}, 395 (1982);

  M.~Gross,
  Phys.\ Lett.\  B {\bf 132}, 125 (1983);

  J.~Bartholomew, D.~Hochberg, P.~H.~Damgaard and M.~Gross,
  Phys.\ Lett.\  B {\bf 133}, 218 (1983).


\bibitem{DKS}
  P.~H.~Damgaard, N.~Kawamoto and K.~Shigemoto,
  Nucl.\ Phys.\  B {\bf 264}, 1 (1986).



\bibitem{Fukushima:2008wg}
  K.~Fukushima,
  Phys.\ Rev.\  D {\bf 77}, 114028 (2008).


\bibitem{Herbst:2010rf}
  T.~K.~Herbst, J.~M.~Pawlowski and B.~J.~Schaefer,
  Phys.\ Lett.\  B {\bf 696}, 58 (2011);
%
  Phys.\ Rev.\ D {\bf 88}, no. 1, 014007 (2013).


\bibitem{Jolicoeur:1983tz}
  T.~Jolicoeur, H.~Kluberg-Stern, M.~Lev, A.~Morel and B.~Petersson,
  Nucl.\ Phys.\  B {\bf 235}, 455 (1984).


\bibitem{deForcrand:2008vr}
  P.~de Forcrand and O.~Philipsen,
  J. High Energy Phys. {\bf 0811}, 012 (2008).

\bibitem{Ichihara:2015bcq} 
  T.~Ichihara, A.~Ohnishi and K.~Morita,
  PoS LATTICE {\bf 2015}, 203 (2016).

\bibitem{Pisarski:1983ms}
  R.~D.~Pisarski and F.~Wilczek,
  Phys.\ Rev.\  D {\bf 29}, 338 (1984).

\bibitem{SCLQCD-OLF}
  I.~Ichinose and K.~Nagao,
  Nucl.\ Phys.\ B {\bf 577}, 279 (2000);
%
  Nucl.\ Phys.\ B {\bf 596}, 231 (2001).

\bibitem{Yu:2005eu} 
  X.~L.~Yu and X.~Q.~Luo,
  Mod.\ Phys.\ Lett.\ A {\bf 22}, 537 (2007).

\bibitem{Langelage}
  G.~Bergner, J.~Langelage and O.~Philipsen,
  J. High Energy Phys. {\bf 1403}, 039 (2014);
%
  J.~Langelage and O.~Philipsen,
  J. High Energy Phys. {\bf 1004}, 055 (2010);
%
  J. High Energy Phys. {\bf 1001}, 089 (2010);
%
  J.~Langelage, G.~Munster and O.~Philipsen,
  J. High Energy Phys. {\bf 0807}, 036 (2008).



\bibitem{Lo:2013etb} 
  P.~M.~Lo, B.~Friman, O.~Kaczmarek, K.~Redlich and C.~Sasaki,
  Phys.\ Rev.\ D {\bf 88}, no. 1, 014506 (2013)
  [arXiv:1306.5094 [hep-lat]].


\bibitem{KlubergStern:1982bs}
  H.~Kluberg-Stern, A.~Morel and B.~Petersson,
  Nucl.\ Phys.\  B {\bf 215}, 527 (1983).

\end{thebibliography}
\end{document}